\documentclass[aps,prb,twocolumn,showpacs,superscriptaddress,floatfix]{revtex4}
\usepackage{graphicx,graphics}
\usepackage{dcolumn}
\usepackage{amsmath,amssymb,amsfonts}
\usepackage{latexsym,verbatim}
\usepackage{bm}
\usepackage{color}
\usepackage{ulem}
\usepackage[breaklinks=true,colorlinks,citecolor=blue,linkcolor=blue,urlcolor=blue]{hyperref}
\newcommand{\bA}{{\bm A}}
\newcommand{\bB}{{\bm B}}

\newcommand{\bp}{{\bm p}}
\newcommand{\br}{{\bm r}}

\newcommand{\vD}{v_{\mathrm{D}}}
\newcommand{\bPi}{{\bm \Pi}}

\begin{document}
\title{Theory of integer quantum Hall polaritons in graphene}
\author{F.M.D. Pellegrino}
\email{francesco.pellegrino@sns.it}
\affiliation{NEST, Scuola Normale Superiore and Istituto Nanoscienze-CNR, I-56126 Pisa, Italy}
\author{L. Chirolli}
\affiliation{Instituto de Ciencia de Materiales de Madrid (CSIC), Sor Juana In\'es de la Cruz 3, E-28049 Madrid, Spain}
\author{Rosario Fazio}
\affiliation{NEST, Scuola Normale Superiore and Istituto Nanoscienze-CNR, I-56126 Pisa, Italy}
\author{V. Giovannetti}
\affiliation{NEST, Scuola Normale Superiore and Istituto Nanoscienze-CNR, I-56126 Pisa, Italy}
\author{Marco Polini}
\affiliation{NEST, Istituto Nanoscienze-CNR and Scuola Normale Superiore, I-56126 Pisa, Italy}
\affiliation{Istituto Italiano di Tecnologia, Graphene Labs, Via Morego 30, I-16163 Genova, Italy}

\begin{abstract}
We present a theory of the cavity quantum electrodynamics of the graphene cyclotron resonance. By employing a canonical transformation, we derive  an effective Hamiltonian for the system comprised of two neighboring Landau levels dressed by the cavity electromagnetic field (integer quantum Hall polaritons). This generalized Dicke Hamiltonian, which contains terms that are quadratic in the electromagnetic field and respects gauge invariance, is then used to calculate thermodynamic properties of the quantum Hall polariton system. Finally, we demonstrate that the generalized Dicke description fails when the graphene sheet is heavily doped, i.e.~when the Landau level spectrum of 2D massless Dirac fermions is approximately harmonic. In this case we ``integrate out'' the Landau levels in valence band and obtain an effective Hamiltonian for the entire stack of Landau levels in conduction band, as dressed by strong light-matter interactions.
\end{abstract}
\maketitle

\section{Introduction}

Light-matter interactions in graphene, a two-dimensional (2D) honeycomb crystal of Carbon atoms~\cite{novoselov_naturemater_2007,castroneto_rmp_2009,Katsnelsonbook}, have been intensively explored in the past decade for both fundamental and applied purposes~\cite{bonaccorso_naturephoton_2010,peres_rmp_2010,koppens_nanolett_2011,grigorenko_naturephoton_2012}.

Recent experimental advances have made it possible to monolithically integrate graphene with optical 
microcavities~\cite{engel_naturecommun_2012,furchi_nanolett_2012}, paving the way for fundamental studies of cavity quantum electrodynamics (QED)~\cite{walter_repprogphys_2006} at the nanometer scale with graphene as an active medium. Another approach, which has been successful~\cite{scalari_science_2012} in achieving the so-called strong-coupling regime of cavity QED~\cite{walter_repprogphys_2006} in conventional 2D electron systems in semiconductor quantum wells, consists in coupling graphene carriers with the photonic modes of an array of split-ring resonators~\cite{valmorra_nanolett_2013}.

Graphene-based cavity QED offers, at least in principle, a number of unique advantages. First, graphene is a highly-tunable active medium since its electrical and heat transport properties can be easily controlled by employing gates~\cite{novoselov_naturemater_2007,castroneto_rmp_2009,Katsnelsonbook}. Second, graphene offers many pathways to achieve the strong-coupling regime: these include i) the exploitation of intrinsic Dirac plasmons~\cite{koppens_nanolett_2011,grigorenko_naturephoton_2012} and ii) the combination of graphene with other plasmonic nanostructures~\cite{echtermeyer_naturecommun_2011}. Third, the active medium can be enriched by employing 2D vertical heterostructures~\cite{verticalheterostructures,novoselov_ps_2012,bonaccorso_matertoday_2012,geim_nature_2013} comprising graphene as well as other 2D crystals/systems such as hexagonal Boron Nitride~\cite{ponomarenko_naturephys_2011,gorbachev_naturephys_2012,britnell_science_2012}, transition metal dichalcogenides~\cite{wang_naturenano_2012,
britnell_science_2013} (e.g.~${\rm MoS}_2$, ${\rm 
WS}_2$, ${\rm WSe}_2$), Gallium Arsenide quantum wells~\cite{principi_prb_2012,gamucci_arXiv_2014}, etc.

A central role in cavity QED is played by the Dicke model~\cite{dicke_pr_1954}, which 
describes a non-dissipative closed system of identical two-level subsystems interacting with a single-mode radiation field. 
For a sufficiently strong light-matter coupling constant, the thermodynamic 
limit of the Dicke model exhibits a second-order quantum phase transition to 
a super-radiant state~\cite{SPT} with macroscopic photon occupation and coherent atomic polarization. 

When an external magnetic field is applied to a 2D electron system, transitions 
between states in full and empty Landau levels (LLs) are dispersionless~\cite{prangebook,allan1994,Giuliani_and_Vignale}, mimicking atomic transitions and enabling~\cite{hagenmuller_prb_2010} a 
condensed matter realization of the Dicke model. 
The light-matter interaction in the Dicke Hamiltonian is linear in the vector potential ${\bm A}_{\rm em}$ of 
the cavity.  For condensed matter states described by parabolic band models,
a quadratic ${\bm A}^2_{\rm em}$ term whose strength is related to the system's 
Drude weight and f-sum rule~\cite{Pines_and_Nozieres,Giuliani_and_Vignale}, 
also emerges naturally from minimal coupling.  It has long been
understood~\cite{nogotheorems} that the Dicke model's super-radiant phase transition is suppressed when the quadratic terms are retained.  Demonstrations of this property are often referred to as {\it no-go} theorems. 

The problem is more subtle in graphene, where electronic states near the charge neutrality point are described in a wide range of energies by a 2D massless Dirac fermion (MDF) Hamiltonian~\cite{Katsnelsonbook,castroneto_rmp_2009}. The MDF Hamiltonian contains only one power of momentum ${\bm p}$: minimal coupling applied to this Hamiltonian does {\it not} generate a term proportional to ${\bm A}^2_{\rm em}$.
The authors of  Ref.~\onlinecite{chirolli_prl_2012} demonstrated that, in the strong coupling regime, the  
model for the cavity QED of the graphene cyclotron resonance 
{\it must} be supplemented by a quadratic term in the cavity photon field that is dynamically generated by inter-band 
transitions and again implies a {\it no-go} theorem. The terms proportional to ${\bm A}^2_{\rm em}$ in the theory of the cavity QED of the graphene cyclotron resonance were derived in Ref.~\onlinecite{chirolli_prl_2012} by using as a guiding principle gauge invariance and by treating inter-band transitions in the framework of second-order perturbation theory.

The main scope of this Article is to lay down a formal theory of the cavity QED of the graphene cyclotron resonance. The key point is that one must derive a low-energy effective Hamiltonian by taking into account the coupling of the two-level systems which are resonant with the cavity photon field to all non-resonant states. This coupling is crucially important in the strong-coupling regime, where all the terms that are proportional to ${\bm A}^2_{\rm em}$, which are generated by our renormalization procedure, must be taken into account. Indeed, these guarantee gauge invariance as well as a {\it no-go} theorem for the occurrence of a super-radiant phase transition, therefore corroborating the findings of Ref.~\onlinecite{chirolli_prl_2012}. Finally, we go beyond this generalized Dicke description by demonstrating that it is not adequate to describe the strong-coupling regime of the cavity QED of the graphene cyclotron resonance in the limit of high doping. In this case we derive and discuss a renormalized 
Hamiltonian for the entire stack of LLs in conduction band, as dressed by the cavity electromagnetic field.

Our Article is organized as following. In Section~\ref{sect:twolevelsystem} we employ a canonical transformation~\cite{hamann_pr_1966,schrieffer_pr_1966,bravyi_annalsofphysics_2011} to derive an effective low-energy Hamiltonian---see Eq.~(\ref{eqn:Heff_sw})---for the cavity QED of the graphene cyclotron resonance and discuss the limits of its validity. We analyze in great detail the invariance of this effective Hamiltonian with respect to gauge transformations by employing a linear-response theory formalism. In Sect.~\ref{sect:thermodynamics} we use a functional-integral formalism to study the thermodynamic properties of the system described by the effective Hamiltonian. In Sect.~\ref{sec:effectivemetalapproach} we transcend the generalized Dicke description of Sect.~\ref{sect:twolevelsystem} and present a renormalized Hamiltonian---see Eq.~(\ref{eqn:Hmetal})---that enables the study of the strong-coupling limit of the cavity QED of the graphene cyclotron resonance in the limit of high doping. Finally, in Sect.~\ref{sect:summary} we report a summary of our main findings and conclusions.

\section{Generalized Dicke Hamiltonian}
\label{sect:twolevelsystem}

In this Section we derive an effective low-energy Hamiltonian for the cavity QED of the graphene cyclotron resonance.

\subsection{Landau levels in graphene}
At low energies, charge carriers in graphene are modeled by the usual single-channel massless Dirac fermion Hamiltonian~\cite{Katsnelsonbook,castroneto_rmp_2009}
 \begin{equation}\label{eqn:weyl}
{\cal H}_{\rm D}=  \vD {\bm \sigma}  \cdot{\bp}~, 
\end{equation}
where $\vD \approx 10^6~{\rm m}/{\rm s}$ is the Dirac velocity. Here ${\bm \sigma} = (\sigma_x,\sigma_y)$ is a 2D vector of Pauli matrices acting on sublattice degrees-of-freedom and ${\bm p} = - i\hbar \nabla_{\bm r}$ is the 2D momentum measured from one of the two corners (valleys) of the Brillouin zone.

A quantizing magnetic field $\bB = B {\hat {\bm z}}$ perpendicular to the graphene sheet is coupled to the electronic degrees-of-freedom by replacing the canonical momentum $\bp$ in Eq.~(\ref{eqn:weyl}) with the kinetic momentum $\bPi= \bp + e \bA_0 /c$, where $\bA_0$ is the vector potential that describes the static magnetic field $\bB$.  The corresponding Hamiltonian is
\begin{equation}\label{eqn:weyl_wB}
{\cal H}_0=  \vD  {\bm \sigma}  \cdot \bPi~.
\end{equation}
We work in the Landau gauge $\bA_0 = - B y {\hat {\bm x}}$. In this gauge the canonical momentum along the ${\hat {\bm x}}$ direction, $p_x$, coincides
with magnetic translation operator~\cite{Giuliani_and_Vignale} along the same direction and it commutes with the Hamiltonian ${\cal H}_0$.  Thus, the eigenvalues of $p_x$ are good quantum numbers.
A complete set of eigenfunctions of the Hamiltonian ${\cal H}_0$ in Eq.~(\ref{eqn:weyl_wB}) is provided by the two component pseudospinors~\cite{goerbig_rmp_2011}
\begin{equation}\label{eqn:LL_rspace}
\langle \br |\lambda, n, k \rangle= \frac{e^{i k x}}{\sqrt{2L}}\left(
 \begin{array}{c}
w_{-, n} \phi_{n-1}(y-\ell_B^2 k)\\
\lambda w_{+, n} \phi_{n}(y-\ell_B^2 k)
\end{array}
\right)~,
\end{equation}
where $\lambda =+1$ ($-1$) denotes conduction (valence) band levels,
$n \in {\mathbb N}$ is the Landau level (LL) index, and $k$ is the eigenvalue of the magnetic translation operator in the ${\hat {\bm x}}$ direction.  In Eq. (\ref{eqn:LL_rspace})
\begin{equation}\label{eq:weights}
w_{\pm, n}=\sqrt{1 \pm \delta_{n, 0}}
\end{equation}
guarantees that the pseudospinor corresponding to the $n=0$ LL has weight only on one sublattice. Furthermore, $\phi_n(y)$ with $n= 0,1,2, \dots$ are the normalized eigenfunctions of a 1D harmonic oscillator with frequency equal to the MDF cyclotron frequency $\omega_{\rm c} =  \sqrt{2} \vD / \ell_B$. Here $\ell_B= \sqrt{\hbar c/(eB)} \simeq 25~{\rm nm}/\sqrt{B[{\rm Tesla}]}$ is the magnetic length. 

The spectrum of the Hamiltonian (\ref{eqn:weyl_wB}) has the well-known form~\cite{goerbig_rmp_2011}
\begin{equation}\label{eqn:LLspectrum}
\varepsilon_{\lambda,n}=\lambda \hbar \omega_{\rm c} \sqrt{n}~.
\end{equation}
Each LL has a degeneracy $\mathcal{N}=N_{\rm f} S /(2 \pi \ell^2_B)$, where $N_{\rm f} = 4$ is the spin-valley degeneracy and $S=L^2$ is the sample area.

\subsection{Total Hamiltonian}

We now couple the 2D electron system described by the Hamiltonian (\ref{eqn:weyl_wB}) to a single photon mode in a cavity. We denote by the symbol 
$\bA_{\rm em}$ the vector potential that describes the cavity photon mode. Carriers in graphene are coupled to the cavity electromagnetic field via the minimal substitution:
\begin{equation}\label{eq:minimalcoupling}
\bPi \to \bPi'= \bPi + \frac{e}{c} \bA_{\rm em}~.
\end{equation}
The cavity vector potential $\bA_{\rm em}$ will be treated within the dipole approximation.
We can neglect the spatial dependence of the electromagnetic field in the cavity
because the photon wavelength is much larger than any other length scale of the system.

Introducing photon annihilation $a$ and creation operators $a^{\dag}$ we can write
\begin{equation}
 \bA_{\mathrm{em}}=\sqrt{\frac{2 \pi  \hbar c^2}{\epsilon \omega V}} {\bm e}_{\mathrm{em}} (a + a^{\dag} ),
\end{equation}
where ${\bm e}_{\rm em}$ is a unit vector describing the polarization of the electromagnetic field, 
 $\omega$ is the photon frequency, $\epsilon$ is the cavity dielectric constant, and $V= L_z L^2$ is 
 the volume of the cavity. Here 
 $L_z \ll L$ is the length of the cavity in the ${\hat {\bm z}}$ direction.

The total Hamiltonian reads 
\begin{equation}\label{eq:Hfull}
{\cal H} = {\cal H}_{\rm em} + {\cal H}_0 + {\cal H}_{\rm int}~,
 \end{equation}
where the first term is the cavity photon Hamiltonian, the second term is the MDF Hamiltonian in the presence of a quantizing magnetic field, i.e.~Eq.~(\ref{eqn:weyl_wB}), 
and the third term describes the coupling between MDFs and the cavity photon mode. More explicitly,
\begin{equation}\label{eq:Hem}
\mathcal{H}_{\rm em}= \hbar \omega \left(a^\dag a + \frac{1}{2} \right)~,
\end{equation}
\begin{equation}\label{eq:H0}
\mathcal{H}_0= \sum_{\lambda,  n,  k} \varepsilon_{\lambda, n} c_{\lambda, n, k}^{\dag} c_{\lambda, n, k}~,
\end{equation}
and
\begin{widetext}
\begin{equation}\label{eq:Hint}
 \mathcal{H}_{\rm int }=
 \frac{g}{\sqrt{\cal N}}
 \sum_{ \lambda,\lambda^\prime , n,  n^\prime, k} 
 \left(\lambda  w_{\lambda,  n} e^-_{\rm em} \delta_{n^\prime, n+1} + \lambda^\prime 
 w_{\lambda^\prime, n^\prime}
  e^+_{\rm em} \delta_{n^\prime, n-1} \right)
\left(a  + a^{\dag}  \right)
c_{\lambda^\prime, n^\prime, k}^{\dag} c_{\lambda, n, k}~.
\end{equation}
\end{widetext}
In Eqs.~(\ref{eq:H0})-(\ref{eq:Hint}) $c^\dag_{\lambda n k}$ ($c_{\lambda, n, k}$) creates (annihilates) an electron with band index $\lambda$, LL index $n$, and wave number $k$. Finally,
\begin{equation}\label{eq:g}
g \equiv \hbar \omega_{\rm c} \sqrt{\frac{e^2}{2 \epsilon L_z \hbar \omega}}~, 
\end{equation}
and $e_{\rm em}^{\pm}=e_{\rm em}^{x}\pm ie_{\rm em}^{y}$, $e_{\rm em}^{x}$ and $e_{\rm em}^{y}$ being the components of the polarization vector ${\bm e}_{\rm em}$.

We consider the integer quantum Hall regime in which a given number of LLs are fully occupied and the remaining ones are empty. Since the MDF Hamiltonian is particle-hole symmetric, we can consider, without loss of generality, the situation in which graphene is $n$-doped and the Fermi energy lies in conduction band ($\lambda=+$).
We denote by $n=M$ the highest occupied  LL. The lowest empty LL is therefore $n=M+1$ and 
the Fermi energy lies in the middle between $n= M$ and $n = M + 1$, i.e.
\begin{equation}\label{eq:Ecalligrafic}
{\cal E}_M \equiv \frac{1}{2}\hbar \omega_{\rm c}(\sqrt{M+1}+\sqrt{M})~.
\end{equation}
\subsection{Canonical transformation}
\label{subsec:SW}

The aim of this Section is to present a systematic procedure that allows us to derive an {\it effective} low-energy Hamiltonian for the LL doublet $n = M, M+1$ as dressed by light-matter interactions. 
We are interested in the case in which the cavity photon is nearly resonant 
with the transition between the two conduction-band LLs $n = M, M+1$:
\begin{equation}\label{eq:resonantcondition}
\hbar\omega \approx \Omega_M \equiv \hbar \omega_{\rm c}(\sqrt{M+1}-\sqrt{M})~. 
\end{equation}
We anticipate~\cite{chirolli_prl_2012} that the effective Hamiltonian will be different from the {\it bare} 
Dicke Hamiltonian that one obtains from Eqs.~(\ref{eq:Hem}), (\ref{eq:H0}), and (\ref{eq:Hint}) by selecting $\lambda = +1$ and $n = M, M+1$, i.e.
\begin{eqnarray}\label{eq:bareDicke}
{\cal H}_{\rm Dicke} &=& {\cal H}_{\rm em} + \sum_{k = 1}^{\cal N}\Bigg[{\cal E}_M \openone_k+
 \frac{\Omega_M}{2}\tau_k^z \nonumber\\
 &+&\frac{g}{\sqrt{\cal N}} (a +a^\dag )(e_{\rm em}^-\tau_k^+ + e_{\rm em}^+\tau_k^-)\Bigg]~.
\end{eqnarray}
Here~\cite{footnotePaulimatrices},  
$\openone_k,\tau^z_k,\tau^\pm_k$ with $k = 1 \dots {\cal N}$ is a set of Pauli matrices that act in the 
$2^{\cal N}$-fold degenerate subspace of the LL doublet $n = M, M+1$,  
$\openone_k$ being the $2\times 2$ identity and $\tau^\pm_k \equiv \left(\tau^x_k \pm i \tau^y_k \right)/2$. More precisely, the final result of the canonical transformation yields a generalized Dicke Hamiltonian of the form---see Eq.~(\ref{eqn:Heff_sw}):
\begin{eqnarray}
{\cal H}_{\rm GDH} &= & {\cal H}_{\rm Dicke} + \Delta_M (a+ a^\dag)^2 \nonumber\\
&+&\sum_{k = 1}^{\cal N}\Bigg[\frac{\kappa}{\cal N}(a +a^\dag )^2 \openone_k -\frac{\kappa^z}{\cal N}(a +a^\dag)^2 \tau_k^z\Bigg]~.\nonumber\\
\end{eqnarray}
We notice that ${\cal H}_{\rm GDH}$ differs from the bare Dicke Hamiltonian (\ref{eq:bareDicke}) because of the presence of three terms that are quadratic in the operator $a + a^\dagger$ and that renormalize both ${\cal H}_{\rm em}$ and the light-matter interaction Hamiltonian. Microscopic expressions for the parameters $\Delta_M$, $\kappa$, and $\kappa^z$ are derived below.

We denote by the symbol ${\cal S}_M$ the subspace of the fermionic Hilbert space spanned by the two LLs which are resonantly coupled to the cavity field, i.e.~$n = M, M+1$, and lay on opposite sides of the Fermi energy. The symbol ${\cal S}_N$, on the other hand, denotes the subspace of the fermionic Hilbert space, which is comprised of all LLs {\it but} $n = M, M+1$. We employ a  canonical transformation with the aim of decoupling the LL doublet $n = M, M+1$ from the ${\cal S}_N$ sector (see Refs.~\onlinecite{hamann_pr_1966,schrieffer_pr_1966,bravyi_annalsofphysics_2011} and also Chapter 8 in Ref.~\onlinecite{Giuliani_and_Vignale}).

Before proceeding further, it is convenient to rewrite the Hamiltonian (\ref{eq:Hfull}) in the following manner:
\begin{equation}\label{eqn:HVV}
{\cal H} ={\cal H}_{\rm em}+{\cal H}_0+V_{\rm D}+V_{\rm O}~,
\end{equation}
where ${\cal H}_{\rm em}$ and ${\cal H}_0$ have been introduced in Eqs.~(\ref{eq:Hem}) and (\ref{eq:H0}), respectively, whereas the light-matter interaction Hamiltonian ${\cal H}_{\rm int}$ has been written as the sum of two terms:  i) $V_{\rm D}$, which connects states either belonging to the subspace ${\cal S}_M$ or 
to the subspace ${\cal S}_N$ and ii)
$V_{\rm O}$, which connects states belonging to different subspaces. 
Therefore $V_{\rm D}$ is a block-diagonal operator with one block referring to the ${\cal S}_M$ subspace and the other one to the ${\cal S}_N$ subspace. In the same representation, ${\cal H}_0$ is trivially a block-diagonal operator since it is a diagonal operator and ${\cal H}_{\rm em}$ is also a block-diagonal operator since it contains only photonic creation and annihilation operators and therefore acts as the identity operator with respect to fermionic labels. On the other hand, $V_{\rm O}$ is a block-off-diagonal operator in the same representation.

We now introduce an unitary transformation
\begin{equation} \label{eqn:Udef}
U=e^{S}~,
\end{equation}
where $S$ is its anti-Hermitian generator. The transformed Hamiltonian reads
\begin{equation}\label{eqn:HSWdef}
{\cal H}^\prime= e^S {\cal H} e^{-S}~.
\end{equation}
The spirit of the canonical transformation~\cite{hamann_pr_1966,schrieffer_pr_1966,bravyi_annalsofphysics_2011} is to transform the original Hamiltonian ${\cal H}$ onto an Hamiltonian ${\cal H}'$ that has no block-off-diagonal terms. A necessary condition to achieve this, is that the generator $S$ be a block-off-diagonal operator. 

The operator $S$ can be found with the desired level of accuracy by following a perturbative approach. We use the Baker-Campbell-Hausdorff formula to rewrite Eq.~(\ref{eqn:HSWdef}):
\begin{equation}\label{eqn:HSW_bch}
{\cal H}' =  \mathcal{H}+ \left[S,\mathcal{H} \right] + 
\frac{1}{2!} \left[S,\left[S,\mathcal{H}  \right] \right]+\ldots~,
\end{equation}
where $[A,B]$ denotes the commutator between the two operators $A$ and $B$. 

We now expand the generator $S$ in a power series:
\begin{equation}
 S=\sum^{\infty}_{j=1} S^{(j)},
 \label{eqn:taylor}
\end{equation}
where $S^{(j)}$ is proportional to $(g_0)^j$, i.e.~the $j$-th power of a suitable dimensionless coupling constant that is controlled by the strength $g$ of light-matter interactions---see Eq.~(\ref{eqn:g_0}) below.

After inserting Eq. (\ref{eqn:taylor}) in Eq. (\ref{eqn:HSW_bch}), we require that each
term of the expansion cancels the corresponding block-off-diagonal term, order by order
in the perturbative expansion in powers of $g_0$. 
This approach leads to a hierarchy of equations, one
for each order in perturbation theory. 

For example, the equation for the generator $S^{(1)}$ up to {\it first} order in $g_0$ reads as follows:
\begin{equation}\label{eqn:diffSW}
[S^{(1)}, {\cal H}_0+{\cal H}_{\rm em}] +  V_{\rm O} = 0~.
\end{equation}
The transformed Hamiltonian is given by the following expression:
\begin{equation} \label{eqn:SW_eff}
{\cal H}' =  {\cal H}_{\rm em}+{\cal H}_0 +  V_{\rm D} + \frac{1}{2}[S^{(1)}, V_{\rm O}] +{\cal O}(g^3_0)~.
\end{equation}
We emphasize that ${\cal H}'$ is correct up to {\it second} order in $g_0$.

\begin{figure}[t]
\begin{center}
\includegraphics[width=\columnwidth]{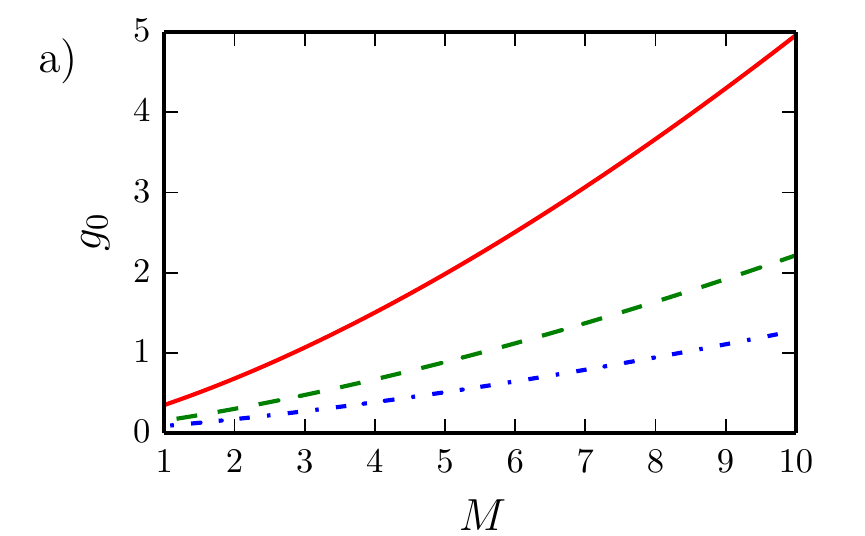}
\includegraphics[width=\columnwidth]{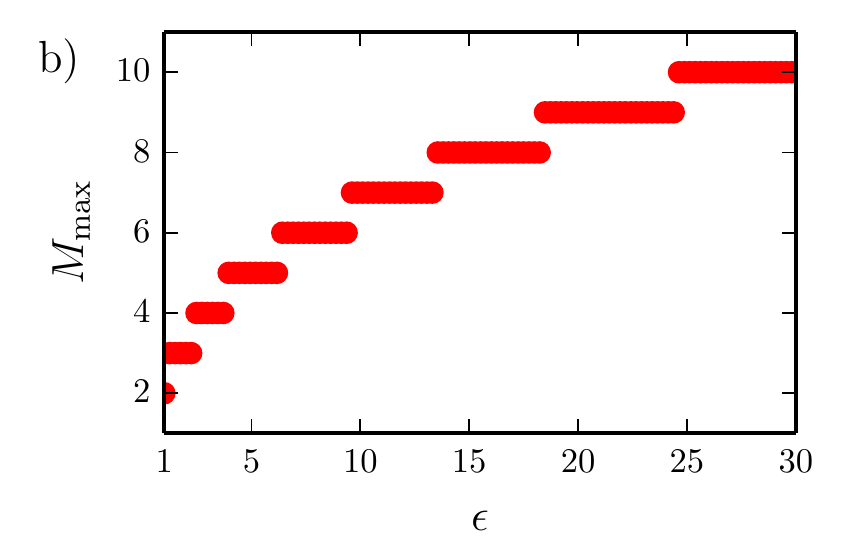}
\end{center}
\caption{Panel a) Dependence of the dimensionless interaction parameter $g_0$, as defined in Eq.~(\ref{eqn:g_0}), on the Landau level index $M$
in the resonant case, i.e.~$\hbar \omega=\Omega_M$. Different curves correspond to different values of the cavity dielectric constant: $\epsilon=1$ (solid line),  
$\epsilon=5$ (dashed line), and $\epsilon=15$ (dash-dotted line). Panel b) Since $g_0$ increases as a function of $M$ for a fixed value of $\epsilon$, we can define the maximum value $M_{\rm max}$ of $M$ up to which $g_0 <1$.    We plot $M_{\rm max}$ as a function of the cavity dielectric constant $\epsilon$. \label{fig:one}}
\end{figure}

The expansion parameter $g_0$ is defined by
\begin{equation}\label{eqn:g_0}
g_0 \equiv \max_{ m \in \mathcal{S}_M, n \in \mathcal{S}_N} {\left(\left|\frac{g}{\hbar \omega -|\varepsilon_{m n}|}\right|\right)}~,
\end{equation}
where $\varepsilon_{m n} \equiv \varepsilon_{m}-\varepsilon_{n}$ 
is the difference between the energies of two LLs. 
From the definition of $g_0$ we clearly see that the canonical transformation cannot be applied if the photon cavity  is resonant with a transition between a LL belonging to the subspace ${\cal S}_M$ and one belonging to the subspace ${\cal S}_N$. As stated above, we are interested in the case in which the cavity photon is nearly resonant 
with the transition between the two LLs in the subspace ${\cal S}_M$, i.e. $\hbar\omega \approx \Omega_M$. Leaving aside the case $M=0$, which needs a separate treatment, the anharmonicity of the LL spectrum in graphene, Eq.~(\ref{eqn:LLspectrum}), 
ensures that the same cavity photon cannot be resonant with other transitions.

In particular, in the resonant case, we obtain $ g_0= g/[\hbar \omega_{\rm c}(\sqrt{M+2}+\sqrt{M}-2 \sqrt{M+1})]$.
If we consider a half-wavelength cavity, we have $\omega=\pi c/(L_z \sqrt{\epsilon})$ and consequently 
$g=\hbar \omega_{\rm c} \sqrt{\alpha/(2 \pi \sqrt{\epsilon})}$, where $\alpha = e^2/(\hbar c) \sim 1/137$ is the QED fine structure constant. Fig.~\ref{fig:one}a) shows a plot of $g_0$ evaluated at $\hbar \omega = \Omega_M$, as a function of the LL index $M$ and for different values of the dielectric constant $\epsilon$. The procedure outlined in this Section is rigorously justified for $g_0<1$. In this regime the LL anharmonicity is larger than the light-matter coupling $g$. Fig.~\ref{fig:one}b) shows that, for a given value of the cavity dielectric constant $\epsilon$, the inequality $g_0 <1$ is satisfied up to maximum value of $M$, denoted by the symbol $M_{\rm max}$, and that one can push the limit of validity of this approach to higher values of $M$ by increasing the value of $\epsilon$.

In Sections~\ref{sec:hamiltonian}-\ref{sect:secondstep} we derive the desired low-energy effective Hamiltonian by using the canonical transformation approach described in this Section. The procedure is carried out in {\it three} steps: i) we first decouple the subspace ${\cal S}_N$ from the subspace ${\cal S}_M$ by applying the canonical transformation $S$ up to first order in the small parameter $g_0$---Eq.~(\ref{eqn:diffSW}); ii) we then use a different canonical transformation to take care of inter-band transitions between LLs belonging to the subspace ${\cal S}_N$; iii) finally, we take into account Pauli blocking.

\subsection{Explicit form of the canonical transformation up to order $g_0$}
\label{sec:hamiltonian}

Following the notation of Sect.~\ref{subsec:SW}, we start from the original Hamiltonian in Eq. (\ref{eqn:HVV}). Here, ${\cal H}_0$, which has been introduced in Eq.~(\ref{eq:H0}), refers to bare electrons in the presence of a quantizing magnetic field and it is diagonal with respect to spin projection, valley index, and the eigenvalue of the magnetic translation operator in the ${\hat {\bm x}}$ direction. It does not couple states belonging to the subspace ${\cal S}_M$ 
with states belonging to the subspace ${\cal S}_N$:
\begin{equation}\label{eqn:sw_H0}
{\cal H}_0= \sum_{m \in {\cal S}_M} \varepsilon_m c_m^\dag c_m + 
\sum_{n \in {\cal S}_N} \varepsilon_n c_n^\dag c_n~.
\end{equation}
Here, $c_m^\dag$ and $c_n^\dag$ ($c_m$ and $c_n$) are fermionic creation (annihilation) operators for a bare electron. 
We emphasize that, in this Section, the indices $m$ and $n$ are {\it collective} labels for the spin projection along the ${\hat {\bm z}}$ axis, 
the valley index, the eigenvalue of the magnetic translation operator in the ${\hat {\bm x}}$ direction, the intra-band LL integer label, and the conduction/valence band label. 

The Hamiltonian that couples electronic degrees-of-freedom with the electromagnetic field is written as a sum of a block-diagonal term $V_{\rm D}$ and a block-off-diagonal term $V_{\rm O}$:
\begin{eqnarray} \label{eqn:sw_VD}
V_{\rm D}&=&\sum_{m,m^\prime \in \mathcal{S}_M} \frac{g_{m m^\prime}}{\sqrt{\mathcal{N}}} \left(a + a^{\dag}  \right) c_m^\dag c_{m^\prime} \nonumber\\
&+& \sum_{n,n^\prime \in \mathcal{S}_N} \frac{g_{n n^\prime}}{\sqrt{\mathcal{N}}}\left(a  + a^{\dag}  \right)c_n^\dag c_{n^\prime}~,
\end{eqnarray}
and
\begin{eqnarray} \label{eqn:sw_VO}
V_{\rm O} &=& \sum_{m \in {\cal S}_M, n \in {\cal S}_N} \Bigg[\frac{g_{m n}}{\sqrt{{\cal N}}}  \left(a  + a^{\dag}  \right) c_m^\dag c_{n} \nonumber\\
&+&\frac{g_{nm}}{\sqrt{{\cal N}}}\left(a  + a^{\dag}  \right)c_n^\dag c_{m}\Bigg]~.
\end{eqnarray}
In Eqs.~(\ref{eqn:sw_VD})-(\ref{eqn:sw_VO}) we have introduced
\begin{equation}
g_{mn}  = \delta_{k, k^\prime}\left(\lambda  w_{\lambda,  {\bar n}} e^-_{\rm em} \delta_{{\bar m}, {\bar n} + 1} + \lambda^\prime 
 w_{\lambda^\prime, {\bar m}}
  e^+_{\rm em} \delta_{{\bar m}, {\bar n} - 1}\right)~,
  \end{equation}
where $n$ ($m$) is the collective label ${\bar n}, \lambda, k$ (${\bar m}, \lambda', k'$). 
Each of these three numbers represents an intra-band LL label (${\bar n}, {\bar m}$), a band index ($\lambda, \lambda'$), and a collective label ($k, k'$) comprising the eigenvalue of the magnetic translation operator in the ${\hat {\bm x}}$ direction, together with the spin projection along the along the ${\hat {\bm z}}$ axis and the valley index. 

By solving  Eq. (\ref{eqn:diffSW}) we obtain an explicit expression for the anti-Hermitian generator $S$ up to first order in $g_0$:
\begin{equation}
S^{(1)} = \sum_{m \in {\cal S}_M, n \in {\cal S}_N} \left(\frac{g_{m n}}{\sqrt{\cal N}} {\cal A}_\omega c_m^\dag c_{n} - \frac{g_{n m}}{\sqrt{\cal N}}c_n^\dag c_{m}{\cal A}^\dagger_{\omega}\right)~,
\end{equation}
where we have introduced the operator
\begin{equation}
{\cal A}_{\omega} \equiv \frac {a}{\varepsilon_{mn}-\hbar \omega} + \frac{a^{\dag} }{{\varepsilon_{mn}+\hbar \omega}}~.
\end{equation}

Given the first-order generator $S^{(1)}$, the commutator $[S^{(1)},V_{\rm O}]$ generates a new block-diagonal term. 
Using the dipole selection rules, the commutator reads
\begin{eqnarray}\label{eqn:comm}
[S^{(1)},V_{\rm O}] & = & 2\left(a  + a^{\dag}  \right)^2\sum_{m \in {\cal S}_M, n \in {\cal S}_N}  \frac{\varepsilon_{mn}}{\varepsilon_{mn}^2-(\hbar \omega)^2} \nonumber\\
&\times&  \frac{g_{mn} g_{nm} }{\mathcal{N}}\left(c_m^\dag c_m - c_n^\dag c_n \right) + {\cal B}_{\omega}~,
\end{eqnarray}
where
\begin{widetext}
\begin{eqnarray}
{\cal B}_{\omega} &=& \frac{2 \hbar \omega}{\mathcal{N}} \left[a,a^\dag \right] \Bigg\{
 \sum_{m \in {\cal S}_M, n \in {\cal S}_N} \frac{g_{mn} g_{nm}}{\varepsilon_{mn}^2
 - \hbar^2 \omega^2}
 \left(c_m^\dag c_m + c_n^\dag c_n \right)+ 
 \sum_{m, m^\prime \in \mathcal{S}_M}  \sum_{n, n^\prime \in \mathcal{S}_N}\Bigg[ 
 \frac{g_{m^\prime n^\prime } g_{mn}}{\varepsilon_{mn}^2- \hbar^2 \omega^2} c_{m^\prime}^\dag c_{n^\prime} c_m^\dag c_n +
\nonumber  \\
&&\frac{g_{ n^\prime m^\prime} g_{mn}}{\varepsilon_{mn}^2- \hbar^2 \omega^2} c_{n^\prime}^\dag c_{m^\prime} c_m^\dag c_n+
 \frac{g_{m^\prime n^\prime } g_{nm}}{\varepsilon_{mn}^2- \hbar^2 \omega^2} c_{m^\prime}^\dag c_{n^\prime} c_n^\dag c_m +
  \frac{g_{ n^\prime m^\prime} g_{n m}}{\varepsilon_{mn}^2- \hbar^2 \omega^2} c_{n^\prime}^\dag c_{m^\prime} c_n^\dag c_m \Bigg]\Bigg \}~.
\end{eqnarray}
\end{widetext}
Note that ${\cal B}_{0} = 0$. More importantly, the operator ${\cal B}_{\omega}$, being proportional to ${\cal N}^{-1}$, is negligible in the limit of a macroscopic LL degeneracy (${\cal N} \gg 1$). 

Using Eq.~(\ref{eqn:SW_eff}) and neglecting terms that are ${\cal O}(g^3_0)$, we finally find the effective Hamiltonian ${\cal H}'$, which is correct up to order $g^2_0$:
\begin{equation}\label{eq:mezzastrada}
{\cal H}' = {\cal H}_{\rm em} + {\cal H}_M + {\cal H}_N~.
\end{equation}
Here, ${\cal H}_M$ is the sum of ${\cal N}$ independent contributions, one for each value 
of $k = 1\dots{\cal N}$, i.e.~${\cal H}_M =\sum^{\cal N}_{k=1} {\cal H}_k$ with
\begin{eqnarray}\label{eqn:H2lv}
{\cal H}_k &=& 
{\cal E}_M \openone_k+
 \frac{\Omega_M}{2}\tau_k^z +\frac{g}{\sqrt{\mathcal{N}}} (a +a^\dag )(e_\mathrm{em}^-\tau_k^+ + e_\mathrm{em}^+\tau_k^-)\nonumber\\
 &-&\frac{\kappa^z}{\mathcal{N}}(a +a^\dag)^2 \tau_k^z+
\frac{\kappa}{\mathcal{N}}(a +a^\dag )^2 \openone_k~,
\end{eqnarray}
where ${\cal E}_M$ and $\Omega_M$ have been introduced earlier in Eqs.~(\ref{eq:Ecalligrafic}) and~(\ref{eq:resonantcondition}), respectively.

The quadratic terms in the electromagnetic field, i.e.~the terms in the second line of Eq.~(\ref{eqn:H2lv}), 
stem from the canonical transformation. In Eq.~(\ref{eqn:H2lv}) we have introduced
\begin{equation}\label{eqn:kz_sd}
\kappa^z \equiv \kappa^z_{\rm s}- \kappa^z_{\rm d}~,
\end{equation} 
where the first term is independent of the cavity photon frequency while 
the second term, that we define ``dynamical'', explicitly depends on the cavity photon frequency:
\begin{equation} \label{eqn:kzs}
\kappa^z_{\rm s}=\frac{g^2}{\Omega_M}
\end{equation}
and
\begin{eqnarray}\label{eqn:kzd}
\kappa^z_{\rm d}&=&
 \frac{\omega^2}{\omega_{\rm c}} \frac{g^2}{\hbar}\Bigg\{\frac{\sqrt{M+1}[\omega^2-(4M+5)\omega^2_{\rm c}]}{[(2 M + 3)\omega^2_{\rm c}-\omega^2]^2 - 4 (M+1)(M+2)\omega^4_{\rm c}}\nonumber\\ 
 &+& \frac{\sqrt{M}[\omega^2-(4M-3)\omega^2_{\rm c}]}{[(2 M  -1)\omega^2_{\rm c} 
 -\omega^2]^2 - 4 M(M-1)\omega^4_{\rm c}}\Bigg\}~.
\end{eqnarray}
Note that $\kappa^z_{\rm d}=0$ for $\omega =0$. Finally,
\begin{eqnarray} \label{eqn:k_omega} 
\kappa &=&\frac{\omega^2}{\omega_{\rm c}}\frac{g^2}{\hbar}\Bigg\{\frac{\sqrt{M+1}[\omega^2-(4M+5)\omega^2_{\rm c}]}{[(2 M + 3)\omega^2_{\rm c}-\omega^2]^2 - 4 (M+1)(M+2)\omega^4_{\rm c}}\nonumber\\
&-&\frac{\sqrt{M}[\omega^2-(4M-3)\omega^2_{\rm c}]}{[(2 M  -1)\omega^2_{\rm c} - \omega^2]^2 - 4 M (M-1)\omega^4_{\rm c}} \nonumber \\
&+&\frac{\sqrt{M+1}-\sqrt{M}}{(\sqrt{M+1}+\sqrt{M})^2\omega^2_{\rm c} - \omega^2}
\Bigg\}~.
\end{eqnarray}
The second term in Eq.~(\ref{eq:mezzastrada}) reads as following:
\begin{widetext}
\begin{equation}\label{eq:HN}
{\cal H}_N=  \sum_{n \in {\cal S}_N} \left[\varepsilon_n +  \sum_{m \in {\cal S}_M}    \frac{\varepsilon_{nm}}{\varepsilon_{nm}^2-(\hbar \omega)^2} 
\left(a  + a^{\dag}  \right)^2 \frac{g_{mn} g_{nm} }{\cal N} \right]c_n^\dag c_n+ \sum_{n,n^\prime \in {\cal S}_N} \frac{g_{n n^\prime}}{\sqrt{\cal N}}\left(a  + a^{\dag}  \right)c_n^\dag c_{n^\prime}~.
\end{equation}
\end{widetext}
\subsection{Elimination of the off-diagonal terms in ${\cal H}_N$ and Pauli blocking}
\label{sect:secondstep}

The Hamiltonian (\ref{eq:mezzastrada}) is not yet the desired result, i.e.~an effective Hamiltonian for the $n=M, M+1$ doublet. Indeed, ${\cal H}_N$ contains fermionic operators that act on the subspace ${\cal S}_N$. 

In particular, we note that the last term in Eq.~(\ref{eq:HN}) is an off-diagonal contribution in the labels $n,n^\prime \in {\cal S}_N$.We utilize a suitable canonical transformation that eliminates this term. For the sake of simplicity, we here report only the final result. We find a renormalized Hamiltonian operating on the subspace ${\cal S}_N$, which is diagonal in the labels $n,n^\prime \in {\cal S}_N$:
 \begin{eqnarray}\label{eqn:HNprime}
{\cal H}^\prime_N &=&  \sum_{n \in \mathcal{S}_N}\varepsilon_n c_n^\dag c_n \nonumber\\
&+&\left(a  + a^{\dag} \right)^2\sum_{n \in \mathcal{S}_N}\sum_{\ell} 
\frac{\varepsilon_{n \ell} g_{n \ell} g_{ \ell n}}{\varepsilon_{n \ell }^2 - \hbar^2 \omega^2} c_n^\dag c_n ~,
 \end{eqnarray}
where the index $\ell$ runs over {\it all} LLs.

Since the Dirac model applies over a large but {\it finite}
energy region, we must regularize~\cite{chirolli_prl_2012} Eq.~(\ref{eqn:HNprime}) by employing a cut-off $\nu_{\rm max}$. Moreover, we treat the fermionic portion of the renormalized Hamiltonian (\ref{eqn:HNprime}) as a mean field for the photons. We therefore replace
\begin{equation}\label{eq:averaging}
c_n ^\dag c_{n} \to  n_{\rm F}(\varepsilon_n) \equiv \frac{1}{\exp{[(\varepsilon_n - \mu_{\rm e})/(k_{\rm B} T)]} +1}~,
\end{equation}
where $\mu_{\rm e}$ is the chemical potential of the electronic system. The accuracy of this mean-field treatment will be justified below in Sect.~\ref{sec:effectivemetalapproach}.

In the low-temperature limit,
\begin{equation}\label{eq:lowtemperaturelimit}
k_{\rm B} T \ll |\varepsilon_n - \mu_{\rm e}|~,~\forall n \in {\cal S}_N~,
\end{equation}
we can replace the Fermi-Dirac function in Eq.~(\ref{eq:averaging}) with a Fermi step. 

We are therefore led to define the prefactor of the $(a + a^\dag)^2$ term in Eq.~(\ref{eqn:HNprime}) as
\begin{equation}
 \Delta_M (\nu_{\rm max})= \sum_{n \in {\cal S}_N}  \sum_{\ell } \frac{\varepsilon_{n \ell}g_{n \ell}g_{ \ell n} }{\varepsilon_{n \ell }^2- \hbar^2 \omega^2} \Theta(\mu_{\rm e} - \varepsilon_n)~, 
\end{equation}
where the sums are regularized with the cut-off $\nu_{\rm max}$.  More explicitly, for every $M \neq 0$, we have:   
\begin{eqnarray}\label{eq:drudeweight}
 \Delta_M (\nu_{\rm max}) &=& -2 \epsilon_{\rm max}\frac{g^2}{\hbar^2 \omega^2_{\rm c}} + 
 \frac{g^2}{\hbar \omega_{\rm c}} {\cal I}_{M-1}  (\nu_{\rm max}) \nonumber\\
&-& \frac{g^2}{\Omega_{M-1}} \frac{\hbar^2 \omega^2}{\Omega^2_{M-1}- \hbar^2 \omega^2}~,
\end{eqnarray}
where $\epsilon_{\rm max} \equiv \hbar \omega_{\rm c} \sqrt{\nu_{\rm max}}$ and
\begin{eqnarray}
 \mathcal{I}_{M-1}(\nu_{\rm max})&=&  
 \sum^{\nu_{\rm max}}_{\ell=M} \Bigg[
 \frac{(\sqrt{\ell+1}-\sqrt{\ell}) \omega^2}{\omega^2- \omega^2_{\rm c} (\sqrt{\ell}+\sqrt{\ell+1})^2}\nonumber\\
 &+& \frac{( \sqrt{\ell}-\sqrt{\ell-1}) \omega^2}
 {\omega^2- \omega^2_{\rm c} (\sqrt{\ell-1}+\sqrt{\ell})^2}\Bigg]~.
\end{eqnarray}
As explained in Refs.~\onlinecite{chirolli_prl_2012,principi_prb_2009}, we must regularize the expression in Eq.~(\ref{eq:drudeweight}) by subtracting the cut-off dependent term $- 2 \epsilon_{\rm max}~g^2/(\hbar^2 \omega^2_{\rm c})$. After applying this regularization, one can take the limit $\nu_{\mathrm{max}}  \to \infty$, discovering that the quantity
\begin{eqnarray}\label{eqn:DMdef}
\Delta_M &\equiv& \lim_{\nu_{\rm max} \to \infty} \left[ \Delta_M(\nu_{\rm max}) + 2 \epsilon_{\rm max}\frac{g^2}{\hbar^2 \omega^2_{\rm c}} \right] \nonumber \\
&=&  \frac{g^2}{\hbar \omega_{\rm c}} {\cal I}^\infty_{M-1} - \frac{g^2}{\Omega_{M-1}} \frac{\omega^2}{\Omega^2_{M-1} - \omega^2}~,
\end{eqnarray}
with
\begin{eqnarray}\label{eq:IinfinityM}
{\cal I}^\infty_M &\equiv & \lim_{\nu_{\rm max} \to \infty} {\cal I}_{M}(\nu_{\rm max}) \nonumber \\
&=& \sum^{\infty}_{\ell=M+1} \Bigg[\frac{ (\sqrt{\ell+1}-\sqrt{\ell}) \omega^2}{\omega^2- \omega^2_{\rm c} (\sqrt{\ell}+\sqrt{\ell+1})^2 }\nonumber\\
&+& \frac{( \sqrt{\ell}-\sqrt{\ell-1}) \omega^2}{\omega^2- \omega^2_{\rm c} (\sqrt{\ell-1}+\sqrt{\ell})^2}\Bigg]~,
\end{eqnarray}
is well defined. 

Discarding constant terms~\cite{Giuliani_and_Vignale} (i.e.~terms that do not contain the photon field operators $a$ and $a^\dag$), the renormalized Hamiltonian (\ref{eqn:HNprime}) becomes
\begin{equation}\label{eq:accaprimofinale}
{\cal H}^\prime_N = \Delta_M \left(a  + a^{\dag} \right)^2~.
\end{equation}
We stress that $\Delta_M$ as defined in Eq.~(\ref{eqn:DMdef}) depends both on the LL label $M$ and the photon frequency $\omega$ and that it vanishes in the static $\omega = 0$ limit.

\subsection{Final result for the effective Hamiltonian}
\label{sec:effective2lv}

In summary, the correct low-energy Hamiltonian is given by ${\cal H}'$ as in Eq.~(\ref{eq:mezzastrada}) with ${\cal H}_N$ replaced by ${\cal H}'_N$ in Eq.~(\ref{eq:accaprimofinale}), i.e.
\begin{equation}\label{eqn:Heff_sw}
{\cal H}_{\rm GDH} \equiv  {\cal H}_{\rm em} +\Delta_M(a +a^\dag )^2+\sum^{\cal N}_{k=1}{\cal H}_k~,
\end{equation}
where ${\cal H}_k$ has been defined in Eq.~(\ref{eqn:H2lv}) and, without loss of generality, we have chosen a specific polarization of the electromagnetic field, {\it i.e.} ${\bm e}_{\rm em}={\bm u}_x$. 

Eq.~(\ref{eqn:Heff_sw}) is the first important result of this Article and represents a low-energy effective Hamiltonian for the cavity QED of the graphene cyclotron resonance. It is evident that ${\cal H}_{\rm GDH}$ differs from the bare Dicke Hamiltonian (\ref{eq:bareDicke}) since it contains terms that are quadratic in the electromagnetic field. We will therefore refer to the low-energy effective Hamiltonian (\ref{eqn:Heff_sw}) as to generalized Dicke Hamiltonian (GDH).

As discussed earlier and as illustrated in Fig.~\ref{fig:one}b), the GDH (\ref{eqn:Heff_sw}) is rigorously justified only for a finite interval of values of $M$, which depends on the cavity dielectric constant. For example, for $\epsilon=15$, Eq.~(\ref{eqn:Heff_sw}) is justified in the interval $0< M \leq 8$. This implies that for this value of $\epsilon$ the description of the cavity QED of the graphene cyclotron resonance in terms of the GDH may break down for $M \geq 9$. Below we discuss an alternative approach, which is valid for arbitrarily large values of the highest-occupied LL index $M$ and transcends the description based on the GDH.

For future purposes, it is useful to highlight the following identity,
\begin{equation}\label{eqn:DM}
\Delta_M =  \frac{g^2}{\Omega_M}+\frac{g^2}{\hbar \omega_{\rm c}}{\cal I}^\infty_{M}-\kappa^z-\kappa~, 
\end{equation}
and the following inequality
\begin{equation}\label{eqn:cfr}
F_M(\omega) \le {\cal I}^\infty_{M} \le  F_{M+1}(\omega)~,
\end{equation}
which is valid $\omega \le \omega_{\rm c} \sqrt{M}$. Here
\begin{equation}\label{eq:deffunction}
F_M(\omega) \equiv \frac{\omega}{2\omega_{\rm c}} \log{\left(\frac{2 \omega_{\rm c} \sqrt{M}-\omega}{2 \omega_{\rm c} \sqrt{M}+\omega}\right)}
\end{equation}
For large $M$ one therefore finds
\begin{equation}
{\cal I}^\infty_{M} \simeq   \frac{\omega}{2\omega_{\rm c}} \log{\left(\frac{2 \mathcal{E}_M-\hbar \omega}{2 \mathcal{E}_M + \hbar\omega}\right)}~.
\end{equation}

In the resonant $\hbar \omega=\Omega_M$ case, the quantities $\kappa^{z}$ and $\kappa$ defined earlier in Eqs.~(\ref{eqn:kz_sd})-(\ref{eqn:k_omega}) reduce to:
\begin{equation}
\kappa^{z}=-\frac{g^2}{\hbar \omega_{\rm c}}\frac{1}{2}\sqrt{M}~,
\end{equation}
and
\begin{eqnarray}\label{eqn:kdyn_res}
\kappa &=&
\frac{g^2}{\hbar \omega_{\rm c}}\Bigg[(M+1)\sqrt{M+1}+\left(M-\frac{1}{2} \right)\sqrt{M} \nonumber\\
&+& 
\frac{1}{4\sqrt{M(M+1)}(\sqrt{M+1}+\sqrt {M})^3} \Bigg]~.
\end{eqnarray}
\subsection{Linear-response theory analysis}
\label{sect:linearresponsetheory}

In this Section we demonstrate that the GDH (\ref{eqn:Heff_sw}) is gauge invariant. 

To this end, we treat the cavity electromagnetic field as a weak perturbation with respect to the MDF Hamiltonian in the presence of a quantizing magnetic field. The cavity electromagnetic field induces a matter current that can be calculated by the powerful means of linear response theory~\cite{Giuliani_and_Vignale,Pines_and_Nozieres}. In particular, the physical matter current in response to the electromagnetic field is composed by paramagnetic and diamagnetic contributions~\cite{Giuliani_and_Vignale,Pines_and_Nozieres}.

It is easy to demonstrate that the paramagnetic response function of a system described by the GDH (\ref{eqn:Heff_sw}) to the electromagnetic field is given by
\begin{eqnarray} \label{eqn:chi_para}
 \chi_{\rm P}(\omega) &=& \frac{g^2}{\mathcal{N}} \langle \langle \tau_{\mathrm{tot}}^x; \tau_{\mathrm{tot}}^x \rangle \rangle_{\omega}\nonumber \\
 &=& g^2 \frac{2 \Omega_M}{\hbar^2 \omega^2 - \Omega^2_M} 
 \tanh{\left(\frac{\beta \Omega_M}{4}\right)}~,
\end{eqnarray}
where $\tau_{\mathrm{tot}}^x=\sum_{k=1}^{\cal N} \tau_{k}^x$ and $\beta=1/(k_{\rm B} T)$. In Eq.~(\ref{eqn:chi_para}) we have introduced the Kubo product~\cite{Giuliani_and_Vignale}
\begin{equation}\label{eq:Kubo}
\langle \langle A; B \rangle \rangle_{\omega} \equiv -\frac{i}{\hbar}\int_0^{\infty}dt~e^{i (\omega + i 0^+)t}\langle[A(t),B]\rangle~,
\end{equation}
where $\langle \dots \rangle$ denotes a thermal average and $A(t)$ is the operator $A$ in the Heisenberg representation, i.e.~$A(t) \equiv \exp(i {\cal H}_{\rm GDH} t)A\exp(-i {\cal H}_{\rm GDH} t)$.

Similarly, the diamagnetic response function is given by
\begin{eqnarray}\label{eqn:chi_dia}
 \chi_{\rm D}(\omega) &=&
 \frac{2}{\mathcal{N}} \langle \langle \kappa \openone_{\mathrm{tot}}-\kappa^z \tau_{\mathrm{tot}}^z \rangle \rangle_{\omega}+  2\Delta_M  \nonumber\\
 &=& 2 \kappa+  2\Delta_M +  2 \kappa^z  \tanh{\left(\frac{\beta\Omega_M}{4}  \right)}~,
\end{eqnarray}
where $\tau^z_{\rm tot}=\sum_{k=1}^{\cal N}\tau^z_k$ and $\openone_{\rm tot}
=\sum_{k=1}^{\cal N}\openone_{k}$. 

The diamagnetic response function $ \chi_{\rm D}(\omega)$ can be rewritten in a compact form as
\begin{equation}\label{eq:Omg}
\chi_{\rm D}(\omega)=2 \Omega_g~,
\end{equation}
where
\begin{eqnarray} \label{eq:Omg_beta}
\Omega_g=  \Omega_g(\beta)&\equiv& \frac{g^2}{\Omega_M} +\frac{g^2}{\hbar \omega_{\rm c}} {\cal I}^{\infty}_M  \nonumber \\
   &-& \kappa^z \left[1- \tanh \left( \beta \Omega_M/4 \right)\right]~.
\end{eqnarray}
In writing Eqs.~(\ref{eq:Omg})-(\ref{eq:Omg_beta}) we have used the mathematical identity (\ref{eqn:DM}).

Therefore, the physical current-current response function is the sum of these two contributions:
\begin{eqnarray}\label{eqn:chi_j}
\chi_{\rm J}(\omega) &=& \chi_{\rm P}(\omega) + \chi_{\rm D}(\omega) \nonumber \\
&= &g^2 
\frac{2 \Omega_M}{\hbar^2\omega^2-\Omega^2_M}\tanh{\left(\frac{\beta \Omega_M}{4}\right)} +2 \Omega_{g}~.  
\end{eqnarray}

In the static $\omega=0$ limit we have
\begin{equation}\label{eq:omegatozero_1}
\chi_{\rm P}(\omega \to 0 )= -\frac{2g^2}{\Omega_M}\tanh{\left(\frac{\beta \Omega_M}{4}\right)}
\end{equation}
and
\begin{eqnarray}\label{eq:omegatozero_2}
\chi_{\rm D}(\omega\to0)&=& 2 \kappa^z_{\rm s} \tanh{\left(\frac{\beta \Omega_M}{4}\right)} \nonumber \\
&=& \frac{2g^2}{\Omega_M}\tanh{\left(\frac{\beta \Omega_M}{4}\right)}~.
\end{eqnarray}
Paramagnetic and diamagnetic contributions in Eqs.~(\ref{eq:omegatozero_1})-(\ref{eq:omegatozero_2}) are
equal in magnitude and opposite in sign. Hence, a quasi-homogeneous vector potential does not induce any response in the static limit:  in this limit the vector potential represents a pure gauge and cannot induce any physical effect unless gauge invariance is broken~\cite{Pines_and_Nozieres,Giuliani_and_Vignale}.

Alert readers will note that the paramagnetic contribution to the physical current-current response function dominates over the diamagnetic contribution in the resonant limit $\hbar\omega \to \Omega_M$. Indeed, $\chi_{\rm P}(\omega)$ has a pole at $\hbar\omega \to \Omega_M$, while $\chi_{\rm D}(\omega)$ is finite at the same frequency. As we will see below in Sect.~\ref{sect:thermodynamics}, however, the quadratic terms in the photon field in Eq.~(\ref{eqn:Heff_sw}), which yield a finite diamagnetic response, are absolutely crucial to ensure thermodynamic stability of the system.

In passing, we notice that the current-current response function in Eq.~(\ref{eqn:chi_j}) has the following asymptotic behavior
\begin{equation}\label{eq:asymptotics}
\chi_{\rm J}(\omega) \to \frac{g^2}{\hbar^2 \omega^2_{\rm c}} \left[2 {\cal E}_M  + \frac{\hbar \omega}{2}  
 \log{\left(\frac{2 {\cal E}_M - \hbar \omega}{2 {\cal E}_M + \hbar\omega}\right)}\right]~,
\end{equation}
in the limit of zero temperature and for $M$ such that $\Omega_M \ll \hbar \omega < 2{\cal E}_M$.
Eq.~(\ref{eq:asymptotics}) is formally identical to the current-current response function of a doped graphene sheet  
in the absence of a quantizing magnetic field~\cite{principi_prb_2009}, provided that one replaces ${\cal E}_M$ 
with the Fermi energy $\mu_{\rm e}$.

 \subsection{Comparison with the findings of Ref.~\onlinecite{chirolli_prl_2012}} 

For the sake of completeness, we now compare the main result obtained so far, i.e.~the GDH (\ref{eqn:Heff_sw}), with the results of Ref.~\onlinecite{chirolli_prl_2012}. 

We start by recalling the effective Hamiltonian 
that was derived  in Ref.~\onlinecite{chirolli_prl_2012}. In the notation of this Article, it reads
\begin{eqnarray}\label{eqn:Hdicke_prl}
{\cal H}_{\rm eff}&=& \hbar \omega \left(a^\dag a + \frac{1}{2} \right) + \sum^{\cal N}_{k=1}\left[{\cal E}_M \openone_k+
 \frac{\Omega_M}{2}\tau^z_k\right.\nonumber \\
&+&\left.\frac{g}{\sqrt{\cal N}} (a +a^\dag ) \tau^x_k -
\frac{\kappa^z_{\rm s}}{\cal N} (a +a^\dag )^2 \tau^z_k\right]~.
\end{eqnarray}
Note that the term proportional to $(a +a^\dag )^2$ in the previous equation contains the Pauli matrix $- \tau^z_k$: this corrects a mistake that was made in Ref.~\onlinecite{chirolli_prl_2012}.

Although the Hamiltonian (\ref{eqn:Hdicke_prl}) respects gauge invariance in the sense of Sect.~\ref{sect:linearresponsetheory}, it misses dynamical contributions that are naturally captured by the canonical transformation. The GDH Hamiltonian (\ref{eqn:Heff_sw}), indeed, reduces to Eq.~(\ref{eqn:Hdicke_prl}) when the dynamical contributions $\kappa^z_{\rm d}$, $\kappa$, and $\Delta_M$ are set to zero. We remind the reader that in the static $\omega \to 0$ limit $\kappa^z_{\rm d}, \kappa, \Delta_M \to 0$.

\section{Thermodynamics of the GDH}
\label{sect:thermodynamics}

In this Section we present a thorough analysis of the thermodynamic properties of the 
GDH (\ref{eqn:Heff_sw}).

The partition function ${\cal Z}$ in the grand-canonical ensemble reads
\begin{equation}
 {\cal Z}={\rm Tr}\left[e^{-\beta ({\cal H}_{\rm eff} - \mu_{\rm ph} N_{\rm ph} - \mu_{\rm e}N_{\rm e})}\right]~,
\end{equation}
where  $N_{\rm ph}$ ($N_{\rm e}$) is the photon (electron) number and $\mu_{\rm ph}$ ($\mu_{\rm e}$) 
is the chemical potential of the photonic (electronic) system. 
Here, we assume that the chemical potential of the electronic system is fixed at ${\cal E}_M$, while the chemical potential of the photons is set to zero.

In order to evaluate the grand-canonical partition function we use the functional integral formalism~\cite{negeleorlandbook}. In this formalism the grand-canonical partition function ${\cal Z}$ is written as a functional integral over bosonic and Grassmann fields:
\begin{eqnarray}\label{eqn:Zgc}
{\cal Z} &=& \int {\cal D}[\phi^\ast(\tau),\phi(\tau)]  \int {\cal D}[\xi_{j k}^\ast(\tau),\xi_{j k}(\tau)] \nonumber\\
&\times& e^{-\mathcal{S}[\phi^\ast(\tau),\phi(\tau),\xi_{j k}^\ast(\tau),\xi_{j k}(\tau)]}~.
\end{eqnarray}
Here, $\phi^\ast(\tau),\phi(\tau)$ represent bosonic fields, which are defined on the imaginary-time interval $[0,\beta]$ and repeated periodically elsewhere, whereas
$\xi_{j k}(\tau),\xi_{j k}^\ast(\tau)$ are Grassmann fermionic fields, which are anti-periodic in the same imaginary-time interval. In Eq.~(\ref{eqn:Zgc}) $k=1,\dots,{\cal N}$ and $j$ labels the eigenvalues of the $2 \times 2$ matrix $\tau^z$, i.e. $j = \pm 1$. Finally, the Euclidean action ${\cal S}$ reads
\begin{widetext}
\begin{eqnarray}\label{eqn:eucl}
 {\cal S}&=& \int^{\beta}_0 d \tau~\left\{\phi^\ast(\tau)\left( \frac{\partial}{\partial \tau}+\hbar \omega \right) \phi(\tau)
 + \Delta_M \left[\phi^\ast(\tau)+\phi(\tau)\right]^2\right\}+
 \sum_{k, j, j^\prime} \int^{\beta}_0 d \tau~\xi_{j k }^\ast(\tau) \left(
\frac{\partial}{\partial \tau} {\openone}_{j j^\prime}+
\frac{\Omega_M}{2} \tau^z_{j j^\prime} \right) \xi_{j^\prime k}(\tau) +
 \nonumber \\
 &&
 \sum_{k, j, j^\prime} \int^{\beta}_0 d \tau~\xi_{j k }^\ast(\tau) \left\{\frac{g}{\sqrt{\mathcal{N}}}\left[\phi^\ast(\tau)+\phi(\tau)\right] \tau^x_{j j^\prime}
 +\left[\phi^\ast(\tau)+\phi(\tau)\right]^2
 \left (\frac{\kappa}{\mathcal{N}} \delta_{j j^\prime}-\frac{\kappa^z}{\mathcal{N}}\tau^z_{j j^\prime} \right)  \right\}\xi_{ j^\prime k }(\tau)~.
\end{eqnarray}
\end{widetext}
\subsection{Static path approximation}
\label{sect:SPA}

The simplest approximation to evaluate the grand-canonical partition function ${\cal Z}$ in Eq. (\ref{eqn:Zgc}) is the so-called ``static path approximation'' (SPA). In the SPA the dependence
of the bosonic fields $\phi^\ast(\tau), \phi(\tau)$ on imaginary time is neglected. 
Therefore, quantum fluctuations of the electromagnetic field are absent in the SPA. 
The SPA is a good approximation when the average photon number is macroscopic, i.e.~when it is ${\cal O}({\cal N})$.
This is precisely what occurs in a super-radiant phase.

The gran-canonical partition function in the SPA reads
\begin{eqnarray}\label{eqn:Z_spa}
{\cal Z}_{\rm SPA} &\equiv& \int \frac{ d \phi^\ast d \phi}{2 \pi i}  \int {\cal D}[\xi_{j k}^\ast(\tau),\xi_{ j k}(\tau)] \nonumber\\
&\times&e^{-{\cal S}[\phi^\ast, \phi, \xi_{j k}^\ast(\tau), \xi_{j k}(\tau)]}~,
\end{eqnarray}
where $\phi^\ast$ and $\phi$ are just complex numbers and not fluctuating fields. 

Carrying out the integral over the Grassmann fields $\xi_{j k}^\ast(\tau),\xi_{ j k}(\tau)$ and over $\Im m(\phi)$, we find
\begin{equation}\label{eqn:Z_sd_gc}
{\cal Z}_{\rm SPA}=\sqrt{ \frac{ {\cal N}}{ \pi \beta \hbar \omega}}  \int^\infty_0  dx~e^{{\cal N } \Phi(x)}~,
\end{equation}
where $x=\Re e(\phi)/\sqrt{\cal N}$ and
\begin{widetext}
 \begin{equation}\label{eqn:func_gc_inf}
 \Phi(x)=-\beta 
(\hbar \omega+   4\Delta_M +4\kappa) x^2 
  +\log
  \left[
    2 \cosh{\left( \frac{\beta  \Omega_M}{2} \sqrt{\left(1-\frac{8 \kappa^z }{\Omega_M} x^2\right)^2+\frac{16 g^2 }{\Omega^2_M} x^2}    
 \right ) 
 +2 \cosh{ \left( 4 \beta \kappa x^2  \right) }
 }
  \right]~.
\end{equation}
\end{widetext}
In the limit ${\cal N} \gg 1$ the integral in Eq.~(\ref{eqn:Z_sd_gc}) can be calculated by employing the steepest descent method~\cite{negeleorlandbook}, i.e.
\begin{equation}\label{eq:steepestdescent}
{\cal Z}_{\rm SPA}\simeq \sqrt{ \frac{2}{\beta \hbar \omega  |\Phi^{\prime \prime}(x_0)|}}
e^{ \mathcal{N}   \Phi(x_0)}~.
\end{equation}
Here $x_0$ denotes a maximum, i.e.
\begin{equation}\label{eq:saddlepoint}
\Phi^{\prime}(x_0) \equiv \left.\frac{d\Phi(x)}{dx}\right|_{x = x_0} = 0
\end{equation}
and
\begin{equation}
\Phi^{\prime \prime}(x_0) \equiv \left.\frac{d^2 \Phi(x)}{dx^2}\right|_{x = x_0} < 0~.
\end{equation}
We now look for solutions of the saddle-point equation (\ref{eq:saddlepoint}).

\begin{figure}[t]
\centering
\includegraphics[width=\columnwidth]{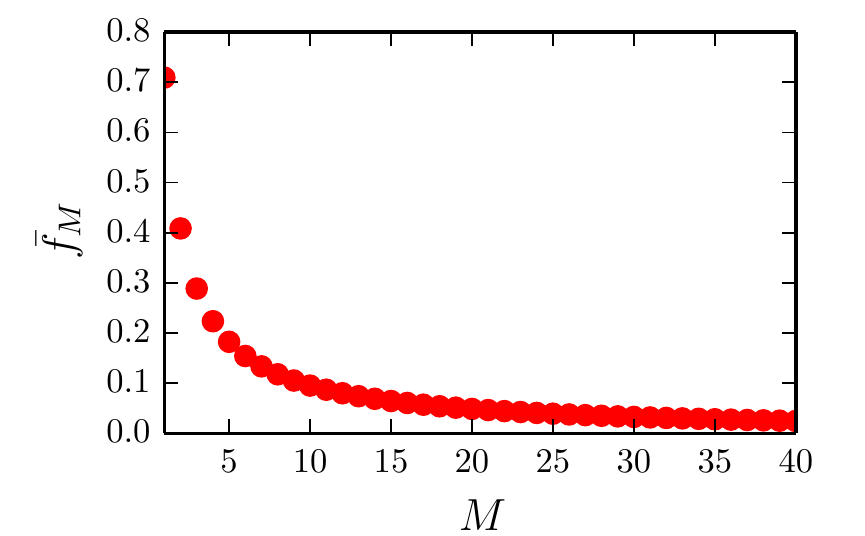}
\caption{Dependence of the function ${\bar f}_M$ defined in Eq.~(\ref{eqn:barfm}) on the LL index $M$.\label{fig:two}}
\end{figure}

Since $\Phi(x)$ depends on $x$ through $x^2$---see Eq.~(\ref{eqn:func_gc_inf})--- 
$x_0=0$ is always an extremum of $\Phi(x)$.
Physically, the solution $x_0=0$ corresponds to the ``normal phase'' in which the 
number of photons vanishes in the thermodynamic limit. We study the nature of this extremum by evaluating $\Phi^{\prime \prime}(0)$. Straightforward algebraic manipulations yield
\begin{eqnarray}\label{eq:phipp0}
\Phi^{\prime \prime}(0)&=&-2\beta \left\{\hbar \omega+4 g^2 {\cal I}^{\infty}_M/(\hbar \omega_{\rm c}) \right.  \nonumber \\
&+&\left.4(g^2/\Omega_M - \kappa^z)[1-\tanh(\beta \Omega_M/4)]\right\}~.
\end{eqnarray}
Since $g^2/\Omega_M>\kappa^z$ and $g/(\hbar\omega_{\rm c})<1$, 
the quantity $\Phi^{\prime \prime}(0)$ can satisfy $\Phi^{\prime \prime}(0) \geq 0$ if and only if the dimensionless function
\begin{equation} \label{eqn:barfm}
{\bar f}_M(\omega) \equiv -4\frac{\omega_{\rm c}}{\omega}{\cal I}^{\infty}_M~,
\end{equation}
is {\it larger} than unity. Note that ${\bar f}_M(\omega)$ is {\it independent} of the cavity dielectric constant $\epsilon$. Since we are interested in the resonant regime, we can set $\omega=\Omega_M/\hbar$ in Eq.~(\ref{eqn:barfm}).  In this case ${\bar f}_M$ becomes a function of the LL label $M$ only. 
Fig.~\ref{fig:two} illustrates the dependence of ${\bar f}_M = {\bar f}_M(\omega=\Omega_M/\hbar)$ 
on $M$. We clearly see that ${\bar f}_M(\omega=\Omega_M/\hbar)<1$ for every $M$. 
We can therefore conclude that $x_0 = 0$ is always a maximum, i.e.~$\Phi^{\prime \prime}(0)<0$.

In what follows, we investigate the possibility of having a super-radiant phase, i.e.~a phase with a macroscopic number of photons in the thermodynamic limit. 
This phase corresponds to the existence of a maximum of $\Phi(x)$ located at a non-zero value of the order parameter $x$. We will show that if $g_0 < 1$ no such extremum exists. This implies that the GDH (\ref{eqn:Heff_sw}) is {\it not} unstable towards a super-radiant state in the regime where its derivation based on the canonical transformation (Sect.~\ref{sect:twolevelsystem}) is rigorously justified.

\subsubsection{Absence of a super-radiant phase}
\label{sect:proofPhi}

We now prove that the saddle-point equation (\ref{eq:saddlepoint}) does not admit any solution at $x_0 \neq 0$. To this end, we write $\exp[{\cal N} \Phi(x)]$ as a sum of functions which are all {\it concave downwards} and have a maximum at $x_0 = 0$.  This is easily accomplished by exploiting the binomial theorem:
\begin{equation}\label{eq:binomialtheorem}
(A + B)^n = \sum_{m = 0}^n 
\left(
\begin{array}{c}
n\\
m
\end{array}
\right)A^{n - m} B^{m}~.
\end{equation}
Using Eq.~(\ref{eq:binomialtheorem}) in Eq.~(\ref{eqn:Z_sd_gc}) we find
\begin{equation}\label{eqn:bin} 
e^{ {\cal N}\Phi(x)} = \sum^{{\cal N}}_{\ell=0} \binom{{\cal N}}{\ell} e^{ {\cal N}   \Phi_\ell (x)}~,
\end{equation}
where
\begin{eqnarray}\label{eqn:Fell}
\Phi_\ell (x) &=& \log(2) -\beta 
(\hbar \omega +  4 \Delta_M+ 4\kappa) x^2 \nonumber\\
&+& \frac{{\cal N} - \ell}{\cal N}\log
 \left [
 \cosh{ \left( 4 \beta \kappa x^2  \right) }
 \right]\nonumber\\
& +&
 \frac{\ell}{\cal N}
 \log
 \left[ \cosh{\left( \frac{\beta  \Omega_M}{2} \chi(x)    
 \right ) }
\right]~,
\end{eqnarray}
with
\begin{equation}
\chi(x) \equiv \sqrt{\left(1-\frac{8  \kappa^z}{\Omega_M}   x^2\right)^2+\frac{16 g^2 }{\Omega^2_M} x^2}~.
\end{equation}
Let us now study the solutions of the equation
\begin{equation}\label{eqn:nullification}
\frac{d \Phi_\ell(x)}{dx} = 0~.
\end{equation}
We first notice that Eq.~(\ref{eqn:nullification}) admits always the trivial solution $x=0$ because $\Phi_\ell(x)$ depends on $x$ only through $x^2$. We now investigate whether solutions exist at non-zero values of $x$. The trivial $x=0$ solution can be easily discarded by taking the first derivative of  $\Phi_\ell(x)$ with respect to $x^2$. Requiring that this vanishes is equivalent to finding the solutions of the following equation:
\begin{eqnarray}\label{eqn:SRP_Fl}
\hbar \omega &+& 4 \Delta_M+4 \kappa\left[1-  \frac{{\cal N}-\ell}{\cal N} \tanh\left(4 \beta \kappa x^2\right)\right] \nonumber\\
&=& \frac{\ell}{\cal N}\frac{\Omega_M}{2}\tanh\left[{\frac{\beta \Omega_M}{2}\chi(x)}\right] \frac{d \chi(x)}{d (x^2)}~.
\end{eqnarray}
Eq.~(\ref{eqn:SRP_Fl}) can also be written as following:
\begin{equation}\label{eqn:biq_ell}
c^{(\ell)}_4(x) x^4 +  c^{(\ell)}_2(x) x^2 +  c^{(\ell)}_0(x) = 0~,
\end{equation}
where
\begin{eqnarray}
c^{(\ell)}_0(x) &=& [\hbar \omega+4\Delta_M+4 \kappa t^{(\ell)}_1(x)]^2\nonumber\\
&-& \left[ \frac{4 \ell}{\cal N} \left(\frac{g^2}{\Omega_M}-\kappa^z\right)
t_2(x)
\right]^2~,
\end{eqnarray}
\begin{eqnarray}
c^{(\ell)}_2(x) &=& \frac{16 \left(g^2/\Omega_M-\kappa^z\right)}{\Omega_M^2}
\Bigg\{\left[\hbar \omega+4\Delta_M+4 \kappa t^{(\ell)}_1(x)\right]^2\nonumber\\
&-& \left[ \frac{4 \ell}{\cal N}  \kappa^z t_2(x)\right]^2 \Bigg\}~,
\end{eqnarray}
and
\begin{eqnarray}
c^{(\ell)}_4(x) &=& \left( \frac{8 \kappa^z}{\Omega_M}\right)^2
\Bigg\{\left[\hbar  \omega+4\Delta_M+4 \kappa t^{(\ell)}_1(x)\right]^2\nonumber\\
&-&\left[ \frac{4 \ell}{\mathcal{N}}  \kappa^z t_2(x)  \right]^2 \Bigg\}~,
\end{eqnarray}
with
\begin{equation}
t^{(\ell)}_1(x) \equiv 1-  \frac{{\cal N} - \ell}{\cal N} \tanh\left( 4 
\beta \kappa x^2\right)
\end{equation}
and
\begin{equation}
t_2(x) \equiv \tanh\left[{ \frac{\beta \Omega_M}{2}\chi(x)}  \right]~.
\end{equation}
From the form of Eq.~(\ref{eqn:biq_ell}) one clearly sees that in order to find a solution of Eq.~(\ref{eqn:nullification}) at finite $x$, one of the coefficients $c^{(\ell)}_4(x)$, $c^{(\ell)}_2(x)$, and $c^{(\ell)}_0(x)$ must change sign for one  value of $\ell$ and $x \neq 0$. 

\begin{figure}[t]
\begin{center}
\includegraphics[width=\columnwidth]{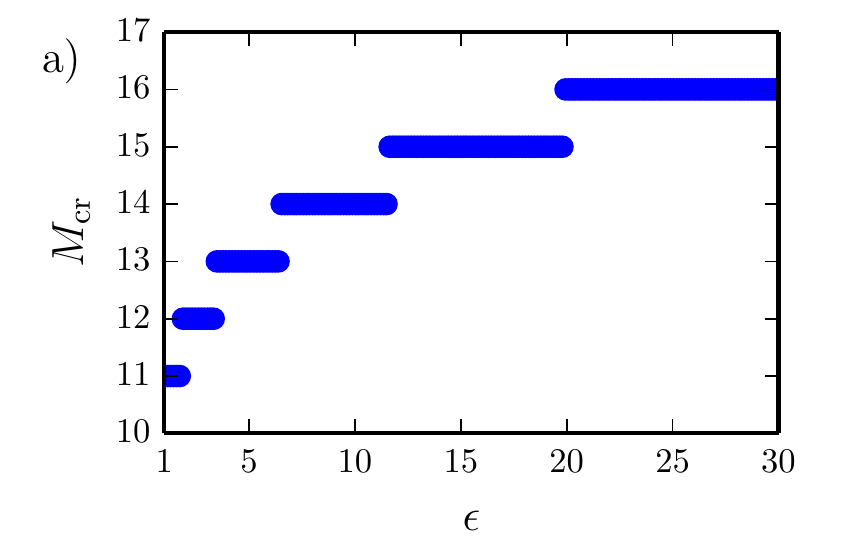}
\includegraphics[width=\columnwidth]{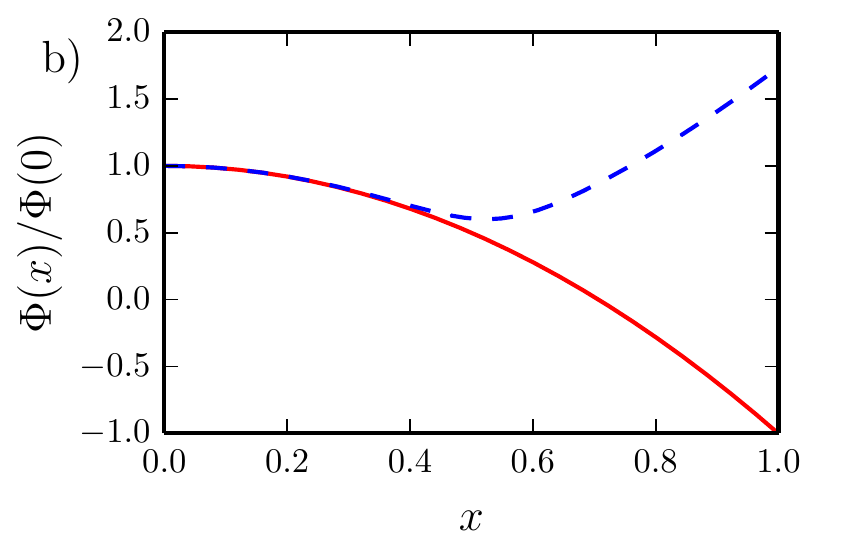}
\end{center}
\caption{Panel a) The quantity $M_{\rm cr}$  is plotted as a function of the cavity dielectric constant $\epsilon$. We remind the reader that 
for $M>M_{\rm cr}$ the condition $f_M (\omega=\Omega_M/\hbar)< 0$ is satisfied. Note that $M_{\rm cr} > M_{\rm max}$---see Fig.~\ref{fig:one}b). Panel b) The ratio $\Phi(x)/\Phi(0)$ as a function of $x$ for $\omega=\Omega_M/\hbar$, $\epsilon=15$, 
and $k_{\rm B} T = 0.1~\Omega_{M=20}$. The solid line refers to $M=5$, which is smaller than the value of $M_{\rm max}$ for $\epsilon = 15$, 
while the dashed line refers to $M =20 \gg M_{\rm cr} > M_{\rm max}$. In this case the GDH (\ref{eqn:Heff_sw}) is not applicable.\label{fig:three}}
\end{figure}

It is easy to see that the functions $c^{(\ell)}_{n}(x)$ with $n = 0, 2$, and $4$ are positive definite for any temperature and any value of $x$ unless the following inequality is satisfied:
\begin{equation}\label{eqn:fM}
f_M(\omega) \equiv \hbar \omega +4 \Delta_ M < 0~.
\end{equation}
Since we are interested in the resonant regime, we can set $\omega=\Omega_M/\hbar$ in Eq.~(\ref{eqn:fM}).  
In this case $f_M$ becomes a function of the LL label $M$ only, i.e.~$f_M = f_M(\omega=\Omega_M/\hbar)$.
We find that, for every value of $\epsilon$, there is a value $M_{\rm cr}$ of the LL index label $M$ 
such that the inequality in Eq.~(\ref{eqn:fM}) is satisfied for $M > M_{\rm cr}$. 
Fig.~\ref{fig:three}a) illustrates the dependence of $M_{\rm cr}$ on $\epsilon$.
By comparison with Fig.~\ref{fig:one}b) we clearly see that $M_{\rm cr} > M_{\rm max}$.
We therefore conclude that the necessary condition for the occurrence of solutions of Eq.~(\ref{eqn:nullification}) at finite $x$, i.e.~$f_M < 0$, cannot be achieved within the limit of validity of the derivation of the GDH (\ref{eqn:Heff_sw}), i.e.~for $M < M_{\rm max}$.

We have therefore demonstrated that, for $M < M_{\rm max}$, $\Phi_\ell (x)$ has no estremum at $x \neq 0$, for every value of $\ell$.
Exploiting the binomial representation in Eq.~(\ref{eqn:bin}), we notice that the function $\exp[{\cal N}\Phi(x)]$ can be written as a sum of concave downwards functions which have a maximum at $x=0$. Therefore $\Phi(x)$
is also concave downwards and has only one maximum at $x=0$.
The function $\Phi(x)$ has neither a global nor a local maximum at $x \neq 0$.  
This implies the impossibility to have a transition to a super-radiant phase.

Fig.~\ref{fig:three}b) shows the quantity $\Phi(x)$ as a function of $x$ for two values of the LL index $M$: $M < M_{\rm max}$ (solid line), where the GDH (\ref{eqn:Heff_sw}) is rigorously justified, and $M \gg M_{\rm cr}$ (dashed line), well beyond the limit of validity of the GDH.
In both cases we see that $\Phi(x)$ has a maximum at $x=0$, as demonstrated earlier. For $M < M_{\rm max}$, no other extremum of $\Phi(x)$ is present. In the case $M \gg M_{\rm crit}$, however, the function $\Phi(x)$ presents a minimum at $x \neq 0$ and 
diverges for $x \gg 1$. More precisely, its is possible to show that $\Phi(x\gg 1) \to -\beta f_M x^2$. 
It follows that the partition function ${\cal Z}_{\rm SPA}$ in Eq.~(\ref{eqn:Z_spa}) is ill-defined for $M \gg M_{\rm cr} > M_{\rm max}$.  
The ``catastrophic'' growth $\Phi(x\gg 1) \to -\beta f_M x^2$ for large $x$ stems from the application of the GDH (\ref{eqn:Heff_sw}) well beyond its limit of validity, i.e.~for $M > M_{\rm crit} > M_{\rm max}$ where $f_M<0$.

Sect.~\ref{sec:effectivemetalapproach} will be devoted to the presentation of a theory that transcends the GDH and that is valid also for  
$M\gg M_{\rm max}$.

\subsubsection{The partition function in the SPA}

We can now finalize the calculation of the partition function in the SPA by following the steepest descent method (\ref{eq:steepestdescent}). We expand $\Phi(x)$ around the maximum at $x=0$ as
\begin{equation}
\Phi(x)\simeq \Phi(0)+\Phi^{\prime \prime}(0)\frac{x^2}{2}~,
\end{equation}
where
\begin{equation}\label{eq:phi0}
\Phi(0)=\log\left[2+2\cosh \left(\beta \Omega_M/2\right)\right]~.
\end{equation}
Using Eqs.~(\ref{eq:phipp0}) and~(\ref{eq:phi0}) in Eq.~(\ref{eq:steepestdescent}), we find
\begin{equation}\label{eqn:zsd_gcan}
{\cal Z}_{\rm SPA} \simeq \frac{{\cal Z}^{(2)}_{\rm free}}{\beta \hbar 
 \omega_g}~,
\end{equation}
where 
\begin{equation}\label{eq:Zfree}
{\cal Z}^{(2)}_{\rm free} \equiv [1+\exp{(\beta \Omega_M/2)}]^{\cal N}[1+\exp{(-\beta \Omega_M/2)}]^{\cal N}
\end{equation}
and
\begin{eqnarray}\label{eq:wg}
\omega_g=\omega_g(\beta)& \equiv& \left\{ \omega  [\omega+4 g^2 {\cal I}^{\infty}_M/(\hbar^2 \omega_{\rm c}) +4(g^2/ \Omega_M - \kappa^z) \right. \nonumber \\ 
&\times& \left.[1-\tanh(\beta \Omega_M/4)]/\hbar ] \right \}^{1/2}~.
\end{eqnarray}
The quantity ${\cal Z}^{(2)}_{\rm free}$ is easily recognized to be the grand-canonical partition function of the LL doublet $n=M, M+1$ in the absence of the cavity photon field.

It is also possible to evaluate the photon occupation number $n^{({\rm SPA})}_{\rm ph}$ in the SPA:
\begin{equation}\label{eqn:spa_n}
n^{({\rm SPA})}_{\rm ph} = -\frac{\partial \log{\cal Z}_{\rm SPA}}{ \partial (\beta \hbar \omega)} =
\frac{1}{\beta \hbar \omega_g}~,
\end{equation}
which is formally identical to the SPA occupation number of a photon gas that does not interact with matter (i.e.~$g=0$),
\begin{equation}\label{eqn:spa_n0}
\left.-\frac{\partial \log{\cal Z}_{\rm SPA}}{ \partial (\beta \hbar \omega)}\right |_{g=0}=\frac{1}{\beta \hbar \omega}~,
\end{equation}
provided that one replaces $\omega \to \omega_g$.
\begin{figure}[t]
\centering
\includegraphics[width=1.00\columnwidth]{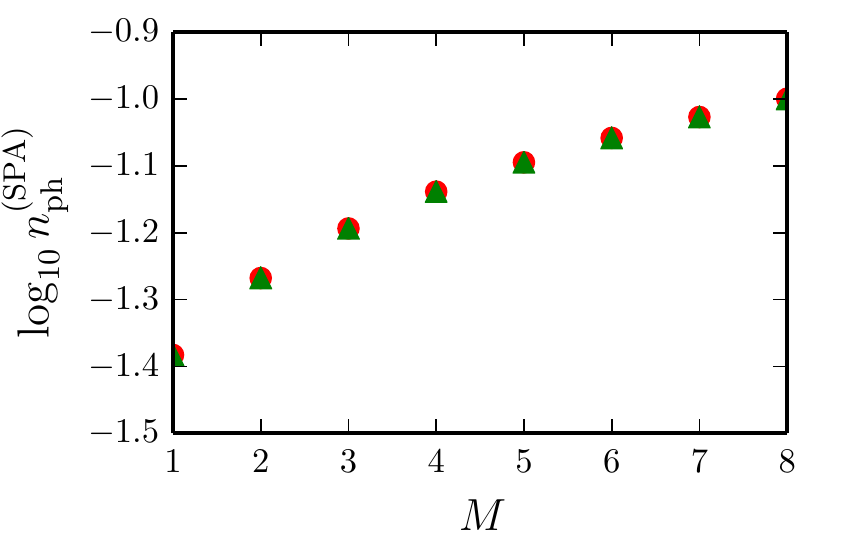}
\caption{The (red) circles denote the logarithm of the photon occupation number $n^{({\rm SPA})}_{\rm ph}$  as a function of the LL label $M$ for $\hbar \omega=\Omega_M$---see Eq.~(\ref{eqn:spa_n}). The (green) triangles denote the SPA photon occupation number evaluated at $g=0$, Eq.~(\ref{eqn:spa_n0}), and for $\hbar \omega=\Omega_M$. In this plot $k_{\rm B} T= 0.1~\Omega_{M = 8}$ and $\epsilon=15$.\label{fig:lognspa}}
\end{figure}
In Fig.~\ref{fig:lognspa} we compare the photon occupation number  $n^{({\rm SPA})}_{\rm ph}$ evaluated on resonance,  $\hbar \omega=\Omega_M$, with the SPA occupation number of the photon gas evaluated at $g=0$, Eq.~(\ref{eqn:spa_n0}). We see that light-matter interactions do not cause any significant modification of the photon occupation number with respect to the $g=0$ case.  We therefore do not see any sign of a super-radiant phase.

\subsubsection{Super-radiance in the absence of the quadratic terms}

We now show that a super-radiant phase transition can occur when the quadratic terms in the photon field are neglected~\cite{SPT}. 

In this case a maximum of $\Phi(x)$ at $x_0 \neq 0$ can occur if~\cite{SPT,hagenmuller_prl_2012}
\begin{equation}\label{eq:nocan}
\frac{\hbar \omega \Omega_M}{4 g^2}<1~. 
\end{equation}
This implies that, choosing a suitable cavity dielectric constant for a given $M$ or a value of the LL index $M$ for a given $\epsilon$, a super-radiant phase transition is possible.
Consider, for instance, a half-wavelength cavity and set $\hbar \omega= \Omega_M$, where 
$\omega=\pi c/(L_z \sqrt{\epsilon})$. In this case $g=\hbar \omega_{\rm c} \sqrt{\alpha/(2 \pi \sqrt{\epsilon})}$ and the critical condition (\ref{eq:nocan}) becomes: $\sqrt{M+1}+\sqrt{M}> 2 \pi \sqrt{\epsilon}/\alpha$. A super-radiant phase transition is therefore possible~\cite{hagenmuller_prl_2012} for large enough values of $M$. 

If the condition (\ref{eq:nocan}) is satisfied, the 
maximum of $\Phi(x)$ appears at
\begin{equation}
x_0=\frac{g}{\hbar \omega}\left[1-\left(\frac{\hbar \omega \Omega_M}{4 g^2} \right)^2\right]^{1/2}
\end{equation}
in the zero-temperature limit. Hence, one can gain energy when the photon occupation number becomes macroscopic, $n^{({\rm SPA})}_{\rm ph} = x_0^2 {\cal N}$.

These are artefacts stemming 
from the neglect of quadratic terms in the photon field.

\subsection{The impact of quantum fluctuations of the electromagnetic field}\label{sect:quantumfluctuations}

Within the SPA, we have demonstrated that the saddle-point equation (\ref{eq:saddlepoint}) admits only the ``trivial'' solution $x=0$, i.e. $\Re e(\phi)=0$, for any value of the temperature $T$. In this Section we present a careful study of the impact of imaginary-time (i.e. quantum) fluctuations of the photonic field $\phi(\tau)$ around $\phi=0$ on the thermodynamic properties of the effective Hamiltonian (\ref{eqn:Heff_sw}). In other words, we want to verify whether the normal phase is robust with respect to quantum fluctuations of the electromagnetic field.
 
We rewrite the Euclidean action ${\cal S}$ in Eq.~(\ref{eqn:eucl}) in the following form:
\begin{widetext}
\begin{eqnarray} \label{eqn:eucl_GSigma}
{\cal S} &=& \int^{\beta}_0 d \tau~\left\{\phi^\ast(\tau)\left( \frac{\partial}{\partial \tau} + 
\hbar \omega \right) \phi(\tau)
 +\Delta_M \left[ \phi^\ast(\tau)+\phi(\tau) \right]^2\right\} \nonumber\\
 &+&
 \sum_{k, j, j^\prime} \int^{\beta}_0 d \tau~\xi_{k j }^\ast(\tau) \left[
-G_0^{-1}(\tau)+ {\Sigma}(\tau) \right]_{j j^\prime} \xi_{k j^\prime }(\tau) ,
\end{eqnarray}
\end{widetext}
where
\begin{equation}
-G_0^{-1}=\frac{\partial}{\partial \tau} \openone+\frac{\Omega_M}{2} \tau^z~,
\end{equation}
\begin{equation}
\Sigma = \Sigma_1+\Sigma_2~,
\end{equation}
\begin{equation}
\Sigma_1 = \frac{g}{\sqrt{\mathcal{N}}}\left[ \phi^{\ast}(\tau)+\phi(\tau) \right]\tau^x~,
\end{equation}
and
\begin{equation}
\Sigma_2 = \left[ \phi^\ast(\tau)+\phi(\tau) \right]^2
\left(\frac{\kappa}{\mathcal{N}} \openone -\frac{\kappa^z}{\mathcal{N}} \tau^z\right)~.
\end{equation}

The key point now is to realize that the fermionic part of the action can be integrated out exactly, since it corresponds to a Gaussian functional integral. The resulting effective action is
\begin{eqnarray}\label{eqn:eucl_eff0}
{\cal S}_{\rm eff} &=& \int_0^{\beta}d \tau~\Bigg\{\phi^\ast(\tau)\left( \frac{\partial}{\partial \tau}
 +\hbar \omega \right) \phi(\tau)\nonumber \\
 &+& \Delta_M \left[ \phi^\ast(\tau)+\phi(\tau) \right]^2\Bigg\} - {\rm Tr}\left[\log \left(-G^{-1}_0 + \Sigma\right)\right]~,\nonumber\\
\end{eqnarray}
where the symbol ``${\rm Tr}$'' means a trace over all degrees-of-freedom, including the imaginary time.

In order to study the effect of Gaussian fluctuations, we expand the last term in the effective action 
${\cal S}_{\rm eff}$ in powers of $\Sigma$ up to {\it second} order in the bosonic fields $\phi^\ast(\tau), \phi(\tau)$. In order to do so, we employ the identity:
\begin{eqnarray}\label{eqn:tr_id}
{\rm Tr}\left[\log \left(-G_0^{-1} + \Sigma  \right)\right] &=&
{\rm Tr}\left[\log \left(-G_0^{-1}  \right)\right] \nonumber \\
&-& {\rm Tr}\sum^{\infty}_{n=1}\frac{(G_0 \Sigma)^n}{n}~.
\end{eqnarray}
Neglecting terms of order $\phi^3(\tau)$ we therefore find:
\begin{eqnarray}\label{eqn:eucl_eff}
{\cal S}_{\rm eff} &\simeq&
\int^{\beta}_0 d \tau~\Bigg\{\phi^\ast(\tau)\left( \frac{\partial}{\partial \tau}+\hbar \omega \right) \phi(\tau) \nonumber\\
&+& \Delta_M \left[ \phi^\ast(\tau)+\phi(\tau) \right]^2 \Bigg\} - {\rm Tr}\left[\log \left(-G_0^{-1}  \right)\right] \nonumber\\ 
&+& {\rm Tr}\left[G_0 \Sigma_2  \right] + \frac{1}{2}{\rm Tr}\left[G_0 \Sigma_1 G_0 \Sigma_1 \right]~.
\end{eqnarray}
We define
\begin{equation}
{\cal S}^{(2)}_{\rm fluct} \equiv {\rm Tr}\left[G_0 \Sigma_2  \right]+ \frac{1}{2}\mathrm{Tr}\left[G_0 \Sigma_1 G_0 \Sigma_1 \right]~.
\end{equation}
The first term in the previous equation is non-zero because $\Sigma_2$ is quadratic in the bosonic fields.

Hence, the grand-canonical partition function in the Gaussian approximation reads
\begin{widetext}
\begin{eqnarray}\label{eq:tobecalculated}
{\cal Z}_{\rm G} &\simeq& {\cal Z}^{(2)}_{\rm free}\int {\cal D}[\phi^\ast(\tau),\phi(\tau)]e^{\displaystyle -\int^{\beta}_0 d \tau~\left\{\phi^\ast(\tau)
\left(\partial/\partial \tau+\hbar \omega \right) \phi(\tau) +\Delta_M \left[ \phi^\ast(\tau)+\phi(\tau) \right]^2
+ {\cal S}^{(2)}_{\rm fluct}(\phi^\ast(\tau),\phi(\tau))\right\}}~.\nonumber\\
\end{eqnarray}
\end{widetext}
where ${\cal Z}^{(2)}_{\rm free}$ has been defined earlier in Eq.~(\ref{eq:Zfree}).
We can now calculate the bosonic functional integral on the right-hand side of Eq.~(\ref{eq:tobecalculated}) since it is a Gaussian functional integral.
This is most easily done by using the Matsubara representation of the photonic field:
\begin{equation}\label{eqn:matsubara}
\phi(\tau)=\frac{1}{\sqrt{\beta}}\sum^{+\infty}_{m=-\infty} e^{- i \omega_m \tau} \phi_m~,
\end{equation}
where $\omega_m=2 \pi m/\beta$ with $m \in {\mathbb N}$. We find
\begin{equation}\label{eqn:Z_matsubara}
{\cal Z}_{\rm G} \simeq {\cal Z}_{\rm SPA}\int \prod^{\infty}_{m=1} \frac{d {\boldsymbol \varphi}_m^\ast
 d {\boldsymbol \varphi}_m}{2 \pi i \beta} e^{-\sum_m {\boldsymbol \varphi}_m^\dag \cdot S_m \cdot {\boldsymbol \varphi}_m}~,
\end{equation}
where ${\boldsymbol \varphi}_m=(\phi_m,\phi_{-m}^\ast)^{\rm T}$ and ${\cal Z}_{\rm SPA}$ has been defined earlier in Eq.~(\ref{eqn:zsd_gcan}).

\begin{figure}[t]
\centering
\includegraphics[width=1.00\columnwidth]{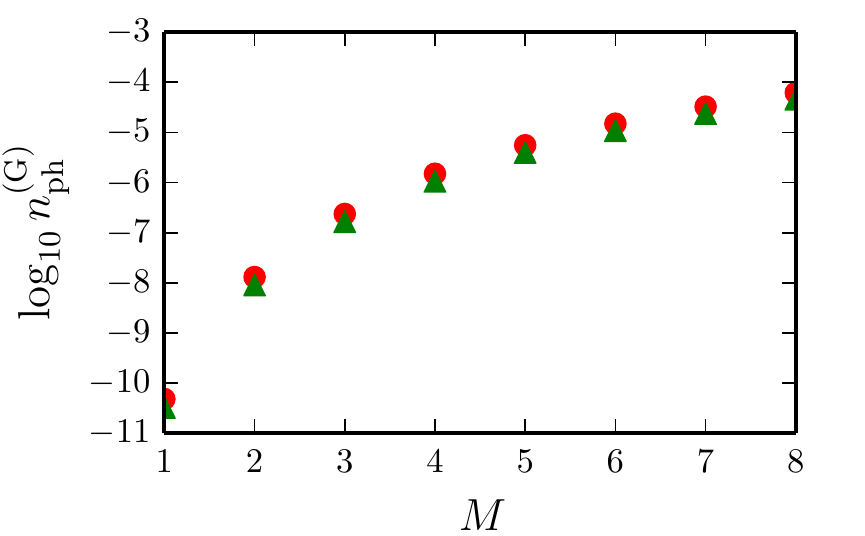}
\caption{The (red) circles denote the logarithm of the photon occupation number $n^{({\rm G})}_{\rm ph}$  as a function of the LL label $M$ for $\hbar \omega=\Omega_M$---see Eq.~(\ref{eqn:n_gauss}). 
The (green) triangles denote the Bose-Einstein thermal factor $n_{\rm B}(\Omega_M)$. 
In this plot $k_{\rm B} T= 0.1~\Omega_{M = 8}$ and $\epsilon=15$.\label{fig:lognb}}
\end{figure}

To evaluate the integral on the right-hand side of Eq.~(\ref{eqn:Z_matsubara}) we need the determinant of the matrix $S_m$. For each positive integer $m$ this reads as follows,
\begin{widetext}
\begin{equation}\label{eqn:det}
{\rm Det}(S_m) =\omega_m^2 + \hbar^2 \omega^2+4 \hbar \omega \left[ \kappa +\Delta_M + \left(\kappa^z -g^2 \frac{\Omega_M}{\omega_m^2+\Omega_M^2} \right)\tanh{\left( \frac{\beta \Omega_M}{4}\right)}\right]~.
\end{equation}
\end{widetext}
We again analyze the resonant case $\hbar \omega=\Omega_M$. It is easy to demonstrate that the function ${\bar f}_M$ in Eq. (\ref{eqn:barfm}) needs to be larger than unity to drive at least one of the determinants $S_m$ to a negative value. But we have already verified that ${\bar f}_M < 1$ for every $M$---see Fig.~\ref{fig:two}. Hence, we have found that the normal phase is robust with respect to quantum fluctuations of the electromagnetic field.

The partition function (\ref{eqn:Z_matsubara}) can be written as
\begin{equation}\label{eqn:Zb_eff}
{\cal Z}_{\rm G} \simeq {\cal Z}_{\rm SPA}\prod^{\infty}_{m=1}\frac{1}{\beta^2 \mathrm{Det}\left( S_m \right)}~.
\end{equation}
We now exploit the identity
\begin{eqnarray}\label{eqn:sum_id}
\frac{1}{\beta \hbar \omega}\prod_{m=1}^{\infty} \frac{1}{\beta^2 (\omega_m^2+ \hbar^2 \omega^2)} &=& \frac{1}{2\sinh(\beta \hbar \omega/2)} \nonumber\\
& \equiv & {\cal Z}_{\rm h.o.}(\omega)~,
\end{eqnarray}
where ${\cal Z}_{\rm h.o.}(\omega)$  is the partition function of an harmonic oscillator with characteristic frequency $\omega$. 

We therefore conclude that the grand-canonical partition function in the Gaussian approximation is given by the following expression:
\begin{equation}\label{eqn:Zfluct}
{\cal Z}_{\rm G} \simeq  {\cal Z}_{\rm SPA}
\frac{(\beta \hbar \omega_+)(\beta\hbar \omega_-)}{\beta \Omega_M} 
\frac{{\cal Z}_{\rm h.o.}(\omega_+){\cal Z}_{\rm h.o.}(\omega_-)}{{\cal Z}_{\rm h. o.}(\Omega_M/\hbar)}~,
\end{equation}
where
\begin{widetext}
\begin{equation}\label{eq:wpm}
 \hbar \omega_{\pm}=\sqrt{\frac{ 
 \hbar\omega \left(\hbar\omega+
 4\Omega_g \right )
+\Omega_M^2
}{2}\pm\sqrt{\frac{
 \left[ \hbar\omega \left(\hbar\omega + 4\Omega_g
 \right) -\Omega_M^2 \right]^2}{4}+4  \hbar  \omega \Omega_M g^2 \tanh\left(\beta \Omega_M/4 \right)}}~.
\end{equation}
\end{widetext}
The quantity $\Omega_g$ has been introduced earlier in Eq.~(\ref{eq:Omg_beta}) and is proportional to diamagnetic response function $\chi_{\rm D}(\omega)$. Physically, the quantities $\omega_{\pm}$ represent the frequencies of the two integer quantum Hall polariton modes. The quantity $\Omega_g$ encodes all the contributions to the polariton modes that stem from quadratic corrections in the photon fields, which are present in the low-energy effective Hamiltonian (\ref{eqn:Heff_sw}). 

Neglecting these terms results in the following integer quantum Hall polariton frequencies~\cite{hagenmuller_prl_2012}:
\begin{widetext}
\begin{equation}\label{eq:wpm_c}
\left. \hbar \omega_{\pm}\right|_{\Omega_g =0} =\sqrt{
 \frac{\hbar^2\omega^2 + \Omega_M^2}{2}\pm
 \sqrt{\frac{\left(\hbar^2\omega^2 
-\Omega_M^2 \right)^2}{4}+4  \hbar  \omega \Omega_M g^2 
\tanh\left(\beta \Omega_M/4 \right)}}~.
\end{equation}
\end{widetext}
\begin{figure}[t]
\centering
\includegraphics[width=1.00\columnwidth]{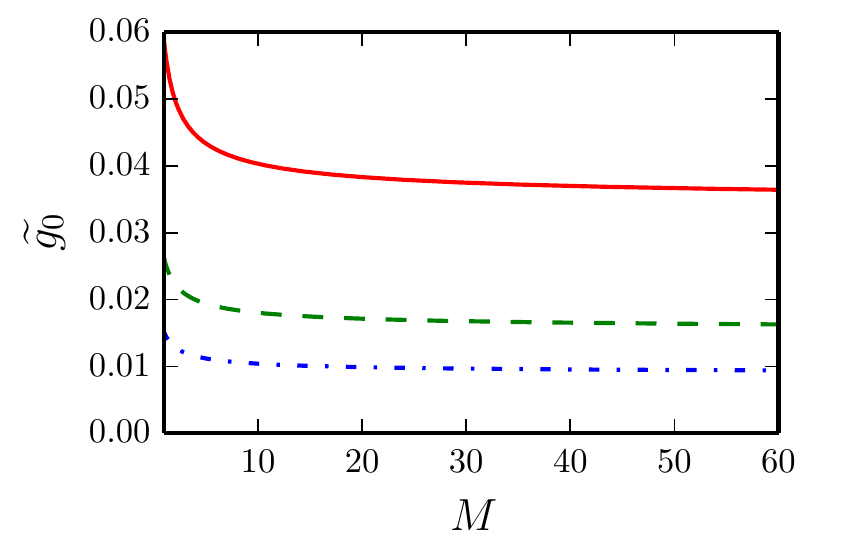}
\caption{Dependence on the LL index $M$ of the smallness parameter ${\widetilde g}_0$ as defined in Eq.~(\ref{eq:smallness_EMM}) and evaluated at $\hbar \omega=\Omega_M$. Different curves correspond to different values of the dielectric constant:
 $\epsilon=1$ (solid line),  $\epsilon=5$ (dashed line), and $\epsilon=15$ (dash-dotted line). \label{fig:met}}
\end{figure}

With the partition function at our disposal, we can evaluate the photon occupation number $n^{({\rm G})}_{\rm ph}$ in the presence of Gaussian fluctuations of the electromagnetic field. We find
\begin{eqnarray}\label{eqn:n_gauss}
n^{({\rm G})}_{\rm ph} &=&-\frac{\partial \log{\cal Z_{\rm G}}}{\partial(\beta \hbar \omega)} = n^{({\rm SPA})}_{\rm ph} \nonumber\\
&+&\sum_{s = \pm}\left[ n_{\rm B}(\omega_s) -  \frac{1}{\beta \hbar \omega_s} \right]  \frac{\partial \omega_s}{\partial \omega}~,
\end{eqnarray}
where $n^{({\rm SPA})}_{\rm ph}$ has been defined in Eq.~(\ref{eqn:spa_n}) and 
$n_{\rm B}(\omega) = [\exp{(\beta \hbar \omega)}-1]^{-1}$ is the Bose-Einstein thermal factor.

In Fig.~\ref{fig:lognb} we compare the photon occupation number $n^{({\rm G})}_{\rm ph}$ evaluated on 
resonance, i.e.~for $\hbar \omega=\Omega_M$, with
the Bose-Einstein function $n_{\rm B}(\Omega_M)$. From this figure we clearly see the photon occupation number obtained from Eq.~(\ref{eqn:n_gauss}) is comparable with the non-interacting photon thermal occupation number. No evidence of a super-radiant phase transition is seen. Comparing $n^{({\rm G})}_{\rm ph}$ in Fig.~\ref{fig:lognb} with $n^{({\rm SPA})}_{\rm ph}$ in Fig.~\ref{fig:lognspa}, we immediately see that the SPA, which treats quasi-classically the electromagnetic field, overestimates the photon occupation number. We have therefore verified that quantum fluctuations of the electromagnetic do not drive the system towards a super-radiant phase and that, on the contrary, suppress the photon occupation number.

\section{Beyond the GDH}
\label{sec:effectivemetalapproach}

As we have discussed above, the description of the cavity QED of the graphene cyclotron resonance in terms of the GDH is not valid for $M \gg M_{\rm max}$, where $M_{\rm max}$ has been illustrated in Fig.~\ref{fig:one}b). In this Section we present a theory that transcends the GDH and that is valid for every $M$.

We again employ a canonical transformation but this time we use it to ``integrate out'' the entire valence band, remaining with an effective Hamiltonian for the entire conduction band as dressed by strong light-matter interactions. With the notation of Sect.~\ref{subsec:SW}, we denote by ${\cal S}_M$ the Hilbert subspace  
spanned by LLs in conduction band, including the zero-energy ($m=0$) LL, whereas  ${\cal S}_N$ denotes the Hilbert subspace spanned by LLs in valence band. In this case the dimensionless parameter that controls the validity of the canonical transformation is
\begin{equation}\label{eq:smallness_EMM}
{\widetilde g}_0 = \frac{g}{|\hbar \omega_{\rm c}-\hbar \omega|}~.
\end{equation}
Fig.~\ref{fig:met} shows ${\widetilde g}_0$ for $\hbar \omega=\Omega_M$ as a function of the LL label $M$.
We clearly see that ${\widetilde g}_0<1$ for any positive $M$ and that ${\widetilde g}_0$ decreases as $M$ increases. Hence, the approach of this Section allows us to study the cavity QED of the graphene cyclotron resonance well beyond the regime of $M$ values where the modeling described in Sect.~\ref{sect:twolevelsystem} works.

Following the approach summarized in Sect.~\ref{subsec:SW}, we find the following effective Hamiltonian for the conduction band:
\begin{widetext}
\begin{eqnarray} \label{eqn:Hmetal}
{\cal H}_{\rm eff} &=& \hbar \omega \left(a^\dag a + \frac{1}{2} \right)+\Delta \left(a+a^\dag \right)^2 \nonumber \\
&+&\sum_{n \in {\mathbb N},k} \left[\varepsilon_{+,n} c_{n, k}^\dag c_{n, k} +  
\frac{w_{+, n} g}{\sqrt{\cal N}}\left(a+a^\dag \right)\left(c_{n, k}^\dag c_{n+1, k} +c_{n+1, k}^\dag c_{n, k} \right)+
\frac{ \kappa_{n}}{\cal N} \left(a+a^\dag \right)^2  c_{n, k}^\dag c_{n, k}\right]~,
\end{eqnarray}
\end{widetext}
where, once again, we have chosen, without loss of generality, a specific polarization of the electromagnetic field,  {\it i.e.} ${\bm e}_{\rm em} = {\bm u}_x$. For the sake of simplicity, we have dropped the label ``$+$'' from the fermionic field operators $c_{+,n,k}$ and $c_{+,n,k}^\dag$. Eq.~(\ref{eqn:Hmetal}) is the second important result of this Article.

In Eq.~(\ref{eqn:Hmetal})
\begin{eqnarray}\label{eq:kappan}
\kappa_n &=& \frac{\left(w_{+, n} g\right)^2 }{\hbar \omega_{\rm c}} 
\frac{(\sqrt{n}+\sqrt{n+ 1}) \omega^2_{\rm c}}{(\sqrt{n}+\sqrt{n+ 1})^2\omega^2_{\rm c} - \omega^2} \nonumber \\
&+& 
\frac{\left(w_{-, n} g\right)^2 }{\hbar \omega_{\rm c}} 
\frac{(\sqrt{n}+\sqrt{n- 1}) \omega^2_{\rm c}}{(\sqrt{n}+\sqrt{n-1})^2\omega^2_{\rm c} - \omega^2}~,
\end{eqnarray}
which is finite in the static $\omega \to 0$ limit, and
\begin{equation}\label{eq:Deltaeffectivemetal}
\Delta=-\frac{g^2}{\hbar \omega_{\rm c}} \frac{\omega^2}{\omega^2_{\rm c}-\omega^2} + \frac{g^2}{\hbar \omega_{\rm c}}{\cal I}^{\infty}_0,
\end{equation}
where ${\cal I}^{\infty}_0$ can be simply obtained by setting $M=0$ in Eq.~(\ref{eq:IinfinityM}). The quantity $\Delta$ in Eq.~(\ref{eq:Deltaeffectivemetal}) vanishes in the static limit. The quantities $w_{\pm, n}$ in Eq.~(\ref{eq:kappan}) have been introduced earlier in Eq.~(\ref{eq:weights}).

\subsection{Thermodynamic properties of the effective Hamiltonian for the entire conduction band: mean-field theory}
\label{sect:MFtheory}

Starting from the effective Hamiltonian in Eq. (\ref{eqn:Hmetal}), we evaluate the grand-canonical partition function ${\cal Z}$ by using again the functional integral formalism. In order to decouple the electronic system from the electromagnetic field, we introduce four complex auxiliary fields, 
i.e.~$y^\ast(\tau),y(\tau)$ and $z^\ast(\tau),z(\tau)$ via the Hubbard-Stratonovich transformation~\cite{negeleorlandbook}:
\begin{widetext}
\begin{eqnarray}\label{eqn:Zgc_HS}
{\cal Z} &=& \int {\cal D}[y^\ast(\tau),y(\tau)] 
\int {\cal D}[z^\ast(\tau),z(\tau)] 
\int {\cal D}[\phi^\ast(\tau),\phi(\tau)]  \int {\cal D}[\xi_{j k}^\ast(\tau),\xi_{j k}(\tau)] \nonumber\\
&\times &\exp{\left[-\sqrt{\cal N} g \int_0^\beta d\tau |y(\tau)|^2 - {\cal N}\frac{g^2}{\hbar \omega_{\rm c}}\int_0^\beta d\tau |z(\tau)|^2 - {\cal S}_{\rm F} 
- {\cal S}_{\rm B}\right]}~,
\end{eqnarray}
where
\begin{subequations}
\begin{eqnarray}\label{eqn:HS}
{\cal S}_{\rm B} &=& \int^{\beta}_0 d \tau~\Bigg\{\phi^\ast(\tau)\left( \frac{\partial}{\partial \tau}+\hbar \omega \right) \phi(\tau)
 +g y(\tau) \left[\phi^\ast(\tau)+\phi(\tau)\right] 
 +\left[\frac{g^2}{\hbar \omega_{\rm c}} z(\tau) + \Delta \right]\left[\phi^\ast(\tau)+\phi(\tau)\right]^2\Bigg\}~,\\
{\cal S}_{\rm F} &=&
 \sum_{k, n } \int^{\beta}_0 d \tau~\Bigg\{\xi_{n k }^\ast(\tau) \left[
 \frac{\partial}{\partial \tau}+ (\varepsilon_{+,n} - {\cal E}_M) - \kappa_n z^\ast(\tau)
 \right] \xi_{n k}(\tau) \nonumber \\ 
 &-&  w_{+, n} g y^\ast(\tau) \left[ \xi_{n k }^\ast(\tau)\xi_{n+1 k }(\tau) + \xi_{n+1 k }^\ast(\tau) \xi_{n k }(\tau)\right]\Bigg\}~.
\end{eqnarray}
\end{subequations}
\end{widetext}
The previous expression for ${\cal Z}$ is formally exact and contains only terms that are quadratic 
in the fermion/boson fields.  In the following, we apply the SPA for the auxiliary complex fields 
by neglecting their imaginary-time dependence and the steepest descent method with respect to the auxiliary fields.
In order to find the saddle point we have to deform
the contours of integration with respect to the static auxiliary fields in the complex
plane~\cite{auerbach_prb_1991}.

 We find that the saddle point is located at:
\begin{subequations}
\begin{eqnarray}\label{eqn:MF}
&&{\bar y}^\ast = - \frac{1}{\sqrt{\cal N}}\langle a +a^{\dag} \rangle_{\rm MF}~,\\
&&{\bar y}=\sum_{k n} \frac{w_{+, n }}{\sqrt{\cal N}} \langle c_{n, k}^\dag c_{n+1, k} +c_{n+1, k}^\dag c_{n, k}	\rangle_{\rm MF}~,\\
&&{\bar z}^\ast= - \frac{1}{\cal N}\langle(a +a^{\dag})^2 \rangle_{\rm MF}~,\\
&&{\bar z}= \frac{\hbar \omega_{\rm c}}{g^2}  \sum_{n, k}  \frac{\kappa_n }{\cal N} 
\langle c_{n, k}^\dag c_{n, k} \rangle_{\rm MF}~,
\end{eqnarray}
\end{subequations}
where the grand-canonical ensemble averages $\langle \ldots \rangle_{\mathrm{MF}}$ are evaluated with respect to the following mean-field Hamiltonian:
\begin{equation}\label{eqn:HMF}
{\cal H}_{\rm MF} = {\cal H}_{\rm B} + {\cal H}_{\rm F}~.
\end{equation}
Here,
\begin{eqnarray}\label{eqn:HMF_B}
{\cal H}_{\rm B} &=& \hbar \omega \left(a^\dag a + \frac{1}{2} \right)+g {\bar y}\left(a+a^\dag \right) \nonumber\\
&+&\left(\frac{g^2}{\hbar \omega_{\rm c}} \bar{z} + \Delta \right)  \left(a+a^\dag \right)^2
\end{eqnarray}
and
\begin{eqnarray}\label{eqn:HMF_F}
 {\cal H}_{\rm F} &=& \sum_{n,k}\Big[\varepsilon_{+, n}c_{n, k}^\dag c_{n, k} 
 -g w_{+, n}{\bar y}^\ast \Big(c_{n, k}^\dag c_{n+1, k} \nonumber \\ 
 &+& c_{n+1, k}^\dag c_{n, k}\Big) - \kappa_n {\bar z}^\ast c_{n, k}^\dag c_{n, k}\Big]~.
\end{eqnarray}
Starting from the bosonic quadratic Hamiltonian ${\cal H}_{\rm B}$, we obtain the following relations between the mean fields:
\begin{subequations}
\begin{eqnarray}
{\bar y}^\ast &=&\frac{\omega g}{\hbar {\bar \omega}^2}  \frac{2 {\bar y}}{\sqrt{\cal N}}~,\label{eqn:y}\\
{\bar z}^\ast&=&-\frac{\omega}{\bar \omega}\frac{2 n_{\rm B}(\hbar {\bar \omega}) +1}{\cal N} - ({\bar y}^\ast)^2~,
\label{eqn:z}
\end{eqnarray}
\end{subequations}
where
\begin{equation}\label{eq:omegabarfunctionofz}
{\bar \omega} = {\bar \omega}({\bar z}) \equiv \sqrt{\omega\left(\omega + \frac{4\Delta}{\hbar} + \frac{4 g^2 {\bar z}}{\hbar^2 \omega_{\rm c}}\right)}~.
\end{equation}
Since the LL degeneracy is macroscopic, i.e.~${\cal N} \gg 1$, in Eq.~(\ref{eqn:z}) we can neglect the first term on the right-hand side and write ${\bar z}^\ast \simeq - ({\bar y}^\ast)^2$. The corresponding mean-field fermionic Hamiltonian (\ref{eqn:HMF_F}) becomes
\begin{eqnarray} \label{eqn:HFy}
 {\cal H}_{\rm F} &\simeq& \sum_{n,k}\Big[\varepsilon_{+, n}c_{n, k}^\dag c_{n, k} 
 -g w_{+, n}{\bar y}^\ast \Big(c_{n, k}^\dag c_{n+1, k} \nonumber \\ 
 &+& c_{n+1, k}^\dag c_{n, k}\Big) + \kappa_n~({\bar y}^\ast)^2~c_{n, k}^\dag c_{n, k}\Big]~.
\end{eqnarray}
For any $\omega>0$, each eigenstate of the mean-field Hamiltonian in Eq. (\ref{eqn:HFy}) has an energy that is a
monotonically increasing function of $|{\bar y}^\ast|$ and has a minimum at ${\bar y}^\ast = 0$. Thus, the self-consistent problem has the following solution:
\begin{subequations}
\begin{eqnarray}
&&{\bar y}^\ast = 0~,\\\label{eqn:MF_sol_1}
&& {\bar y} = 0~,\\\label{eqn:MF_sol_2}
&&{\bar z}^\ast = 0~,\\\label{eqn:MF_sol_3}
&&{\bar z} = \frac{\hbar \omega_{\rm c}}{g^2}  \sum_{n} \kappa_n 
n_{\rm F}(\varepsilon_{+, n})~,\label{eqn:MF_sol_4}
\end{eqnarray}
\end{subequations}
where $n_{\rm F}(x) = [e^{\beta(x - {\cal E}_M)}+1]^{-1}$ is the Fermi-Dirac thermal factor. 
We emphasize that the solution (\ref{eqn:MF_sol_1})-(\ref{eqn:MF_sol_4}) of the mean-field problem posed by the Hamiltonian (\ref{eqn:HMF}) is an {\it a posteriori} check of the mean-field treatment we adopted in Eq.~(\ref{eq:averaging}) of Sect.~\ref{sect:secondstep}.

By using the steepest descent method, we can explicitly write the grand-canonical function as
\begin{equation}\label{eq:MFpartition}
{\cal Z} \simeq {\cal Z}_{\rm MF} \equiv {\cal Z}^{(\infty)}_{\rm free} {\cal Z}_{\rm h.o.} (\bar{\omega})~,
\end{equation}
where ${\cal Z}_{\rm h.o.}(\omega)$ has been introduced in Eq.~(\ref{eqn:sum_id}), 
${\bar \omega}$ is given by Eq.~(\ref{eq:omegabarfunctionofz}) evaluated at ${\bar z}$ as from Eqs.~(\ref{eqn:MF_sol_1})-(\ref{eqn:MF_sol_4}), and
\begin{equation}
{\cal Z}^{(\infty)}_{\rm free} \equiv \prod^{\infty}_{n=0} \left[1+ e^{\beta({\cal E}_M - \varepsilon_{+, n})}\right]^{\cal N}~.
\end{equation}
Note that ${\cal Z}^{(\infty)}_{\rm free}$ is the grand-canonical partition function of the multi-level system $n=0,1,2, \dots$ in the absence of the cavity photon field.

\subsection{Gaussian fluctuations beyond mean-field theory}

In this Section we investigate the stability of the mean-field solution given in the Sect.~\ref{sect:MFtheory} by calculating the fluctuations of the Hubbard-Stratonovich
auxiliary fields~\cite{auerbach_prb_1991}. To this end,  we expand the grand-canonical partition function in Eq.~(\ref{eqn:Zgc_HS}) around its saddle point up to quadratic order. 

Following a procedure analogous to the one sketched in Sect.~\ref{sect:quantumfluctuations}, we find
\begin{equation}
{\cal Z} \simeq \frac{{\cal Z}_{\rm MF}}{\sqrt{D_0}}\prod^{\infty}_{m=1} \frac{1}{D_m }~,
\end{equation}
where
\begin{widetext}
\begin{equation}
D_m =  1-\frac{\omega}{\bar{\omega}}g^2 G_m(\hbar \bar{\omega}) \sum^{\infty}_{n=0} w_{+, n}^2
\left[n_{\rm F}(\varepsilon_{+, n})-n_{\rm F}(\varepsilon_{+, n+1}) \right] G_m(\Omega_n)~,
\end{equation}
\end{widetext}
where ${\bar \omega}$ has been defined after Eq.~(\ref{eq:MFpartition}),  $\Omega_n\equiv \varepsilon_{+, n+1}- \varepsilon_{+, n}$, and $G_m(\Omega)=2 \Omega/[(i \omega_m)^2-\Omega^2]$ with  $\omega_m= 2 \pi m/\beta$.

\begin{figure}[t]
\centering
\includegraphics[width=1.00\columnwidth]{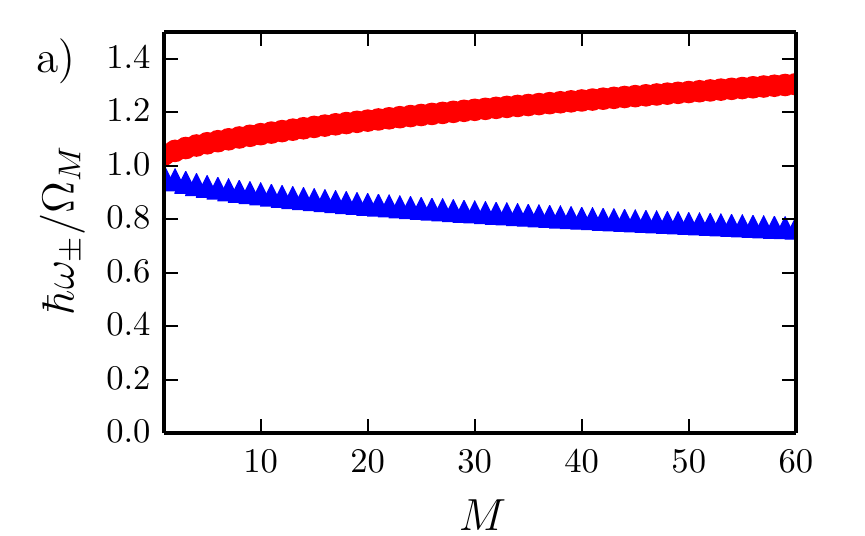}\\
\includegraphics[width=1.00\columnwidth]{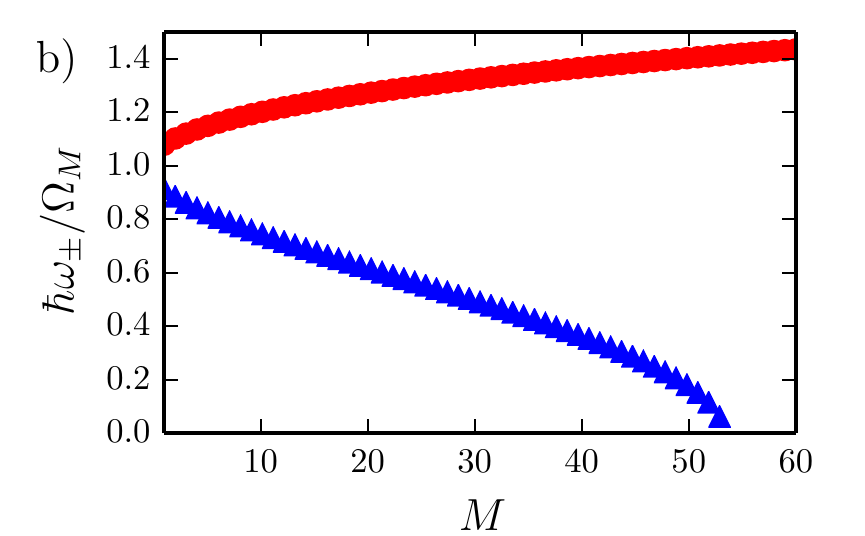}
\caption{Upper and lower integer quantum Hall polariton branches as a function of the LL label $M$. Filled (red) circles denote the upper polariton branch $\hbar \omega_+$ in units of $\Omega_M$ and evaluated on resonance $\hbar \omega=\Omega_M$. Similarly, filled (blue) triangles denote the lower polariton branch $\hbar \omega_-$ in units of $\Omega_M$. In this plot $T = 0$ and $\epsilon=1$. The results in panel a) have been obtained by {\it including} the contribution to the polariton modes that is due to quadratic terms in the electromagnetic field---Eq.~(\ref{eq:wpm}). On the other hand, in panel b) the quantity $\Omega_g$ due to quadratic terms in the electromagnetic field is artificially set to zero---Eq.~(\ref{eq:wpm_c}). In this case the lower polariton branch $\omega_-$ softens at a sufficiently large value of $M$ ($M^\star \simeq 53$ for $\epsilon = 1$) signaling an artificial second-order phase transition to a super-radiant phase.\label{fig:seven}}
\end{figure}

In the low-temperature limit $k_{\rm B} T \ll \Omega_M$ and for $M>0$
\begin{equation}\label{eq:omegabarlowT}
{\bar \omega} \simeq \sqrt{\omega\left[\omega + 4 \left(\frac{g^2}{\hbar \Omega_M} + \frac{g^2}{\hbar^2 \omega_{\rm c}}{\cal I}^\infty_M\right)\right]} 
\end{equation}
and
\begin{equation}\label{eq:detmlowT}
D_m \simeq  1-\frac{\omega}{{\bar \omega}}g^2 G_m(\hbar {\bar \omega}) G_m(\Omega_M)~.
\end{equation}
In writing Eq.~(\ref{eq:detmlowT}) we have used that the Fermi energy lies between the $M$-th and $(M+1)$-th LL, i.e.~that ${\cal E}_M=\hbar \omega_{\rm c} (\sqrt{M+1}+\sqrt{M} )/2$.

It is easy to see that $D_0 > 0$ if and only if 
${\bar f}_M <1$, where ${\bar f}_M$ is defined in Eq. (\ref{eqn:barfm}). This condition has already been discussed in Sect.~\ref{sect:proofPhi} and is always satisfied. Moreover, since $D_m > D_0 $ for any positive integer $m$,  no instability of the mean-field state occurs. Hence, we have demonstrated that the mean-field state is robust with respect to Gaussian fluctuations of
the Hubbard-Stratonovich fields.

The grand-canonical partition function can be written, in the low-temperature limit, as
\begin{widetext}
\begin{equation}\label{eqn:Zgauss}
{\cal Z}=\frac{{\cal Z}^{(\infty)}_{\rm free}}{\beta \hbar \omega_g} 
 \frac{(\beta \hbar \omega_+)(\beta\hbar \omega_-)}{\beta \Omega_M} 
 \frac{\mathcal{Z}_{\mathrm{h. o.}}(\omega_+)\mathcal{Z}_{\mathrm{h. o.}}(\omega_-)}{\mathcal{Z}_{\rm h.o.}(\Omega_M/\hbar)}~,
\end{equation}
\end{widetext}
where $\omega_{\pm}$ are the frequencies of the integer quantum Hall polaritons in the low-temperature limit $\beta \Omega_M \gg 1$---Eq.~(\ref{eq:wpm}) with the replacement $\tanh(\beta\Omega_M/4) \to 1$. Similarly, $\omega_g$ is defined in Eq.~(\ref{eq:wg}) and needs here to be evaluated in the low- temperature limit $\beta \Omega_M \gg 1$, i.e.
\begin{equation}
\omega_g \simeq \sqrt{\omega\left[\omega + 4\frac{g^2}{\hbar^2 \omega_{\rm c}}{\cal I}^\infty_M\right]}~. 
\end{equation}

Figs.~\ref{fig:seven}a)-b) illustrate the dependence of the frequencies $\omega_{\pm}$ on $M$. In particular, panel b) shows that the frequency of the lower polariton branch $\omega_-$ vanishes in the case in which quadratic terms in the photon fields are neglected, i.e. when $\Omega_g$ is set to zero---Eq.~(\ref{eq:wpm_c}). For the values of the parameters chosen in this figure, this occurs at $M^\star \simeq 53$. The softening of the lower polariton branch signals the occurrence of an artificial second-order super-radiant phase transition at a large but finite value of $M$. In Fig.~\ref{fig:seven}a) we see that, for any $M$, the polariton branches $\omega_\pm$ evaluated for $\Omega_g \neq 0$ are positive definite. In particular, Fig.~\ref{fig:seven}a) shows that the frequency of the lower polariton $\omega_-$ is a monotonically decreasing function of $M$: using the definition (\ref{eq:wpm}) we find that $\omega_- \to \Omega_M^2/(2 \hbar g)$ for $M \gg 1$. This result ensures that there is no finite $M$ at 
which 
$\omega_-$ crosses zero, if $\Omega_g \neq 0$. In summary, we have verified that there is no occurrence of super-radiant phase transitions in the cavity QED of the graphene cyclotron resonance. This statement is true also for large values of the highest occupied LL $M$ where the two-level system description adopted in Sect.~\ref{sect:twolevelsystem} fails and one has to resort to the multi-level effective Hamiltonian in Eq.~(\ref{eqn:Hmetal}).

Finally, we highlight that the partition function in Eq.~(\ref{eqn:Zgauss}) formally coincides with the partition function of the two-level system effective model, Eq.~(\ref{eqn:Zfluct}), provided that ${\cal Z}^{(\infty)}_{\rm free}$ is replaced by ${\cal Z}^{(2)}_{\rm free}$.

\section{Summary and conclusions}
\label{sect:summary}

In this Article we have presented a theory of the cavity QED of the graphene cyclotron resonance. 

We have first employed a canonical transformation to derive an effective Hamiltonian for the system comprised of two neighboring Landau levels dressed by the cavity electromagnetic field (integer quantum Hall polaritons). The final result is in Eq.~(\ref{eqn:Heff_sw}). 
This effective Hamiltonian, which we have termed ``generalized Dicke Hamiltonian'', respects gauge invariance and contains terms that are {\it quadratic} in the electromagnetic field. We have then used Eq.~(\ref{eqn:Heff_sw}) and a functional integral formalism to calculate thermodynamic properties of the integer quantum Hall polariton system. We have corroborated the results of Ref.~\onlinecite{chirolli_prl_2012} by confirming that no super-radiant phase transitions are possible in the cavity QED of the graphene cyclotron resonance. 

Starting from a careful analysis of the smallness parameter $g_0$ of the canonical transformation, Eq.~(\ref{eqn:g_0}), we have proved that the generalized Dicke Hamiltonian description fails for sufficiently large value of the highest-occupied Landau level index $M$---see Sect.~\ref{sect:proofPhi}. The maximum value $M_{\rm max}$ of $M$ up to which the derivation of the generalized Dicke Hamiltonian is reliable depends on the value of the cavity dielectric constant $\epsilon$, as illustrated in Fig.~\ref{fig:one}b). For $M > M_{\rm max}$ one has to transcend the generalized Dicke Hamiltonian description. In this case we have used a canonical transformation to project out the entire stack of Landau levels belonging to the valence band. The end result of this approach is an effective Hamiltonian for the entire stack of Landau levels in conduction band, as dressed by light-matter interactions. This result is reported in Eq.~(\ref{eqn:Hmetal}).

In this Article we have discarded electron-electron interactions, which play a very important role in low-dimensional electron systems and, in particular, in the quantum Hall regime where the kinetic energy is quenched and interactions are dominant. Future work will be devoted to understand the role of electron-electron interactions in the theory of quantum Hall polaritons~\cite{pellegrino}.

\acknowledgements
It is a pleasure to thank Allan MacDonald for many enlightening conversations. 
We acknowledge support by the EC under Graphene Flagship (contract no. CNECT-ICT-604391) (M.P.), the European Research Council Advanced Grant (contract no. 290846) (L.C.),  the Italian Ministry of Education, University, and Research (MIUR) through the programs ``FIRB IDEAS" - Project ESQUI (Grant No. RBID08B3FM) (V.G.), ``FIRB - Futuro in Ricerca 2010" - Project PLASMOGRAPH (Grant No. RBFR10M5BT) (M.P.) and PRIN Grant No. 2010LLKJBX (R.F.), and a 2012 SNS Internal Project (V.G.).

\begin{thebibliography}{77}
\bibitem{footnotePaulimatrices}
	By direct comparison of Eqs.~(\ref{eq:Hem})-(\ref{eq:Hint}) with Eq.~(\ref{eq:bareDicke}), 
	we see that the Pauli matrices we have introduced are a shorthand 
	for the following combinations of creation/destruction operators:
%
\begin{subequations}
\begin{eqnarray}
\openone_k&=&c_{+,M+1,k}^\dag c_{+,M+1,k}+c_{+,M,k}^\dag c_{+,M,k}~,\nonumber\\
\tau^z_k&=&c_{+,M+1,k}^\dag c_{+,M+1,k}-c_{+,M,k}^\dag c_{+,M,k}~,\nonumber\\
\tau^+_k&=&c_{+,M+1,k}^\dag c_{+,M,k}~,\nonumber\\
\tau^-_k&=&c_{+,M,k}^\dag c_{+,M+1,k}~.\nonumber 
\end{eqnarray}
\end{subequations}
\end{thebibliography}
\end{document}